# Spin-Spacetime Censorship

Jonathan Nemirovsky[1], Eliahu Cohen[2], Ido Kaminer[1*]

[1]Technion – Israel Institute of Technology, Haifa 32000, Israel

[2]Bar Ilan University, Ramat Gan 5290002, Israel

*Corresponding author: kaminer@technion.ac.il

Quantum *entanglement* and *relativistic causality* are key concepts in theoretical works seeking to unify quantum mechanics and gravity. In this article, we show that the interplay between relativity theory and quantum entanglement has intriguing consequences for the spacetime surrounding elementary particles with spin. Classical and quantum gravity theories predict that a spin-generated magnetic dipole field causes a (slight) bending to the spacetime around particles, breaking its spherical symmetry. Motivated by the apparent break of spherical symmetry, we propose a very general gedanken experiment that does not rely on any specific theory of classical or quantum gravity, and analyze this gedanken experiment in the context of quantum information. We show that any spin-related deviation from spherical symmetry would violate relativistic causality. To avoid the violation of causality, the *measurable* spacetime around the particle's rest frame must remain spherically symmetric, potentially as a back-action by the act of measurement. This way, our gedanken experiment proves that there must be a *censorship mechanism* preventing the possibility of spacetime-based spin detection, which sheds new light on the interface between quantum mechanics and gravity. We emphasize that our proposed gedanken experiment is independent of *any* theory and by allowing spacetime to be quantized its purpose is to be used for testing present and future candidate theories of quantum gravity.

## 1. Introduction

In 1915, general relativity revolutionized our view and understanding of the universe. Black holes, gravity waves, gravitational time dilation, gravitational lensing, and gravitational redshift are just a few amazing examples of its predictive power. Yet even today, and although several promising approaches have been proposed[1-4], we are still unable to reconcile general relativity with the theory of quantum mechanics. This apparent incompatibility has been accentuated by the information paradox in black holes[5] and by the AMPS paradox[6]. These paradoxes, however, do not necessarily indicate whether and how gravity can be unified with quantum mechanics.

The necessity of unifying gravity and quantum mechanics has been the subject of much interest and important debates from Feynman's 1957 gedanken experiment[7], through many analyses over the years[8–15], and even very recently[16–24]. Particularly related to the current paper are works suggesting experiments on the interface between gravity and quantum information[18,19,21,22] that could prove the necessity of quantizing gravity. Such experiments may remove doubts[25,26] regarding the empirical testability of the would-be unified theory. Our aim in this work is different, as we focus on the problems arising when trying to consistently couple quantum spins with spacetime (classical or quantum, we explore both). Our conclusions apply to studies regarding semiclassical gravity, linearized quantum gravity and the ADM formalism, as well as to works in quantum foundations regarding nonlinearity and gravitational decoherence.

Below, we present a new *gedanken* experiment that may shed new light on possible paths towards a unified theory of quantum mechanics and gravity. Specifically, **our gedanken experiment can be used today, without waiting for an experimental realization. It serves** **as a testing ground for the missing theory for quantum measurements of spacetime**, **quantized or not.**

While formulating a unified theory of quantum gravity is a major challenge, even the much simpler question of how to maintain relativistic causality in the quantum world has led to important "no-go" theorems in quantum information. e.g., the "no-signaling" principle[27] or its successors, the "no-communication" theorem[28], which forbids instantaneous transfer of information between two observers, as well as the "no-cloning" theorem[29,30] or the "no teleportation" theorem (see e.g.[31]) which forbid the creation of an identical copy of an arbitrary unknown quantum state. We will show below how such considerations, coming from the field of quantum information, **may give rise to new insights regarding the sought-after description of quantum measurements in quantum gravity.**

Our work presents a gedanken experiment that tests how a spin is coupled to the (either classical or quantum) spacetime around it and what should measurements of the spacetime tell us about the spin. This coupling is measured with clocks that are arranged symmetrically around the spin. The hands of the clock show the time dilation of each clock, which can be used to infer the axis of the spin. Using such a clocks-based spin measurement, we propose a variant of the Einstein-Podolsky-Rosen (EPR) gedanken experiment. To predict the outcomes of such an experiment, one needs a theory that describes the act of measurement in quantum gravity. The paper then continues to discussion parts where we analyze the gedanken experiment with various theories of gravity: classical and quantum. With each candidate theory, we ask whether it maintains relativistic causality in the gedanken experiment, or leads to a paradox.

This simple gedanken experiment potentially has far-reaching implications: either (1) the equations describing the spacetime around the spin must preserve the spherical symmetry of spacetime, in spite of the symmetry being broken by the spin angular momentum and magnetic moment; or (2) the quantum structure of spacetime must not allow ("censor") the determination of the spin state upon measurement. We explore the specific implications of both (1) and (2) on our existing understanding of the interface between quantum mechanics and general relativity. As the first example, we show that if one simply tries to model the gedanken experiment with classical gravity, this requires modifications of Einstein-Maxwell field equations (EMFE) with consequences on astronomical-scale measurements – all arising from the need to maintain relativistic causality. Then we focus on how different theories of quantum gravity maintain causality in our gedanken experiment, and discuss constraints on how to define the act of quantum measurement of spacetime.

The key component in our work is the intrinsic spin of all elementary particles – quantum gravity theories (as well as general relativity) predict that the spacetime and the intrinsic spin are coupled[32]. In particular, the spin is believed to be a source of gravity and is expected to create a minuscule aspherical curvature of spacetime or its quantum analog (by "aspherical" we refer to a broken spherical symmetry). The next section shows how this simple symmetry breaking can be used in our EPR-like gedanken experiment. When analyzing the experiment with certain theories (e.g., classical gravity), the symmetry breaking is found to be the reason for an apparent conflict with relativistic causality, which will have to be carefully circumvented in any self-consistent theory of quantum gravity.

## 2. Presenting the gedanken experiment

We begin by proposing the following gedanken experiment performed in three stages (Fig. 1): At the beginning, **(a)** An EPR pair of entangled electrons, or any spin-½ particles, $\left(\left|\uparrow_A\downarrow_B\right\rangle-\left|\downarrow_A\uparrow_B\right\rangle\right)/\sqrt{2}$ is prepared (this state is basis invariant, of course). One particle is sent

to Alice and the other to Bob (Fig. 1a). **(b)** Alice uses Stern-Gerlach magnets to measure the spin of her electron (Fig 1b). She decides how to orient her magnets, i.e. she can orient them parallel to the $\hat{\mathbf{x}}$ axis or she can orient them parallel to the $\hat{\mathbf{y}}$ axis. Her choice modifies the joint electrons' wavefunction, so that she knows the axis of Bob's spin, which has to be parallel to the axis of her spin (i.e., $|+\hat{\mathbf{x}}\rangle$ or $|-\hat{\mathbf{x}}\rangle$ if her magnet is oriented along $\hat{\mathbf{x}}$, and $|+\hat{\mathbf{y}}\rangle$ or $|-\hat{\mathbf{y}}\rangle$ if her magnet is oriented along $\hat{\mathbf{y}}$). **(c)** Bob places extremely precise clocks at equal distances and different angles around his particle (Fig 1c), and waits long enough to be able to measure and compare minuscule differences in the hands of the various clocks that arise from time dilation. Bob now uses the time dilations in an attempt to determine the spin axis of the electron, i.e., Bob checks whether his spin state is parallel to the $\hat{\mathbf{x}}$ axis or to the $\hat{\mathbf{y}}$ axis.

This way, Bob tries to find how Alice arranged her magnets – whether she chose to orient them parallel to the $\hat{\mathbf{x}}$ axis or parallel to the $\hat{\mathbf{y}}$ axis. If the spacetime curvature around his electron is correlated to its spin axis and thus breaks spherical symmetry, then Bob will be able, in principle, to use the time dilation to determine the spin axis of his electron. This contradicts relativistic causality when Alice is sufficiently far (so that Bob is outside of her light cone).

Clearly, we would not expect violations of causality to be possible. In the rest of the manuscript we shall therefore analyze *how time measurement is described* in different theories of gravity, trying to model the gedanken experiment without contradicting relativistic causality. Therefore, the gedanken experiment serves as a testing ground to examine the underlying assumptions in trying to combine quantum mechanics and general relativity. We will see that **the critical component in the gedanken experiment that leads to this apparent contradiction is the symmetry deviation of the spacetime curvature around the electron being correlated with the electron's spin axis.**

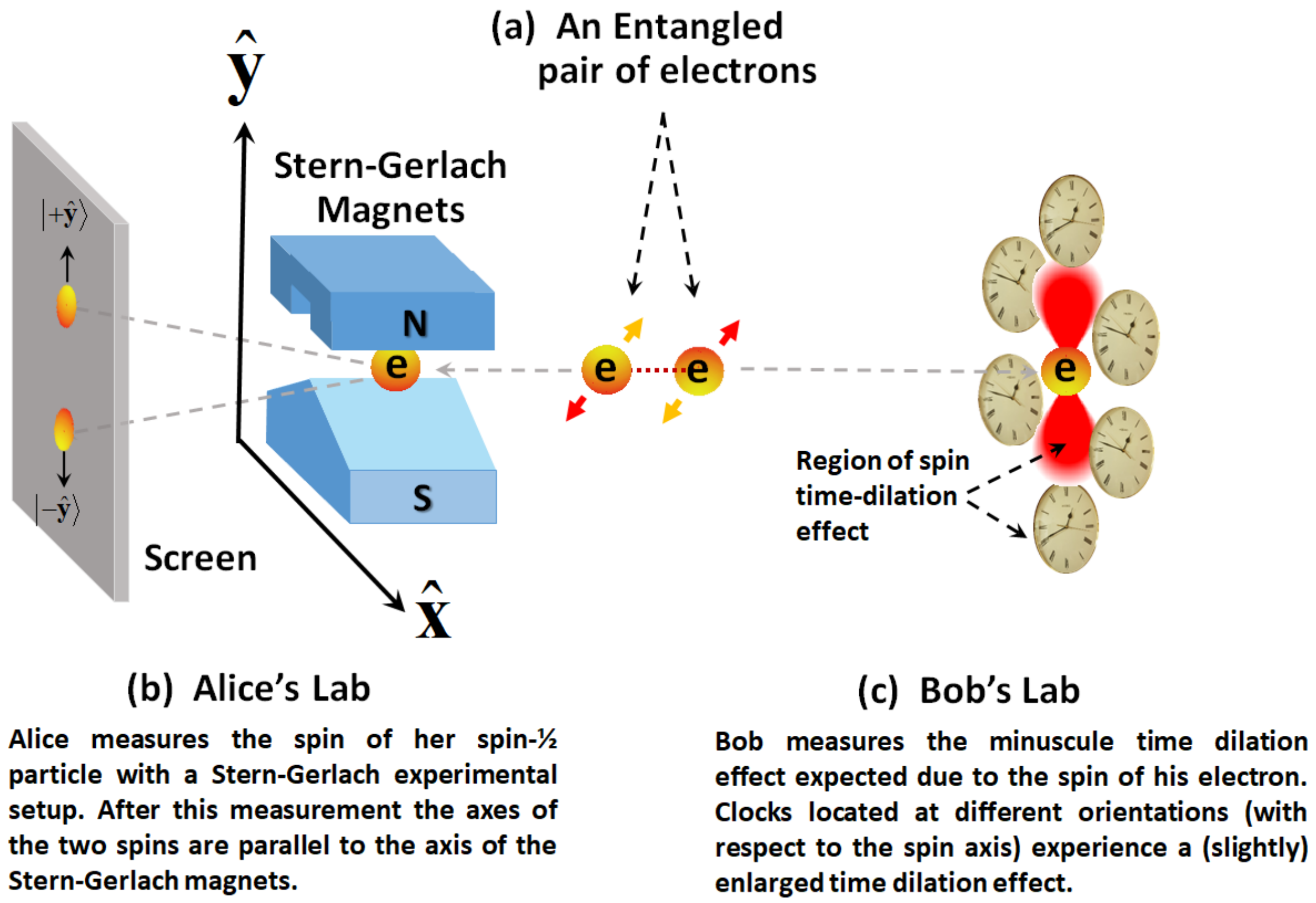

**Figure 1: The gedanken experiment.** The gedanken experiment shows that the measurable spacetime around a spin-½ particle (marked by 'e') cannot show symmetry breaking correlated to the spin axis, or else information can be transmitted at a superluminal speed. The stages of the experiments are as follows: **(a)** A pair of entangled spin-½ particles is prepared and distributed to Alice and Bob, **(b)** Alice performs an ordinary Stern-Gerlach quantum measurement on her particle's spin. She decides to orient her magnets' axes parallel to the $\hat{\mathbf{x}}$ axis or parallel to the $\hat{\mathbf{y}}$ axis. The spin of Bob's particle is then also parallel to the $\hat{\mathbf{x}}$ or $\hat{\mathbf{y}}$ axis (depending on the measurement axis that Alice chose). **(c)** Bob places clocks around the spin and measures the hands of the clocks to infer the time dilation around the particle. Any break of spherical symmetry in the reading of the clocks will allow Bob to determine the axis of the spin and find out how Alice oriented her magnets, even if he is outside of her light cone. This should be impossible because it contradicts causality. **It therefore follows that the *measurable* time dilation must be spherically symmetric, potentially as a back-action by the mere act of measurement, thus *censoring* the spin axis measurement.**

A different description of our gedanken experiment can focus on stage (c) only, whereby Bob can attempt to "clone" the state of a particle using the spacetime measurement, thus violating the "no-cloning" theorem (thus also violating quantum unitarity[33]). Note that this single-particle description of the gedanken experiment has slightly different consequences as it does not require Alice to participate at all, focusing on only one spin state in Bob's lab.

Generally, Bob can make his time measurements arbitrarily precise by accumulating time dilations over prolonged time-like intervals. The effect can also be greatly enhanced by performing the experiment simultaneously with many pairs of entangled particles.

Using the same concepts from quantum information, we propose variants of the gedanken experiment that use clocks to measure other components of spacetime[34] (e.g., elements of $g_{\alpha\beta}$ beyond $g_{00}$). These measurements lead to similar EPR-like tests as they are also expected to be correlated with the axis of the spin (see SI section 1). For another variant of our gedanken experiment that does not use spin at all, see SI section 2.

It is important to emphasize the differences between our gedanken experiment and related EPR experiments that avoid clocks and gravitational effects altogether, measuring instead the electron's spin by its magnetic interaction. In such experiments, the spin state is detected through the magnetic field it creates or by measuring its motion in response to an external magnetic field. In all such cases, the measurement creates a back-action effect that alters the spin state. An inherent difference in our clock-based gedanken experiment is that unlike the magnetic field $\mathbf{B}$ that depends on the spin direction, including its sign, the time dilation effect is (in most theories) independent of the spin sign and only depends on its axis (see Methods section 2 and note e.g., that the energy density $\mathbf{B}^2/2\mu_0$ is sign independent). More generally, spin measurements with magnetic fields cannot reveal with certainty a pre-prepared spin axis or direction because the components of the magnetic field represent quantum operators that do not commute. In contrast, there is no widely accepted way to define a similar commutation relation for the measurement of spacetime nor there is a known gravitational back-action effect from the clocks if they are located symmetrically around the spin. More information appears in Methods section 4 and SI section 1. As another emphasis of the difference, we propose a gedanken experiment entirely *without* the spin in SI section 2, using the same approach from quantum information.

To quantify the gedanken experiment, we employ a density matrix formulation that includes the clocks as part of the quantum system, so that the matrix contains the spin together with the time dilations measured by the clocks (Methods section 1). Employing density matrices can help formalizing the gedanken experiment with multiple candidate theories, as well as linearized quantum gravity. More on that below and in the SI section 8. In particular, further discussion of the back-action on the spin by the act of quantum measurement is in Methods section 4, where we also explore a general framework that allows an entanglement of the spin with the quantized spacetime. As an example, we use the ADM formalism and the resulting Wheeler-DeWitt equation.

### *The necessity of censorship*

The conceptual strength of our gedanken experiment originates from relativistic causality implying that the spacetime around Bob's electron must not indicate any deviation from spherical symmetry *upon measurement* of the time dilation. Any deviation of this sort contains information about Alice's choice of basis that should be precluded on Bob's side, because it reveals the spin axis (without its actual up/down value). This simple result points to a necessary physical mechanism that seems to be missing from the current accepted physical theory: a *spin-censorship mechanism*. This spin-censorship mechanism, which hinders any form of spin detection, is based on the quantum measurement of spacetime, and holds for all the elementary particles (even photons – see SI section 4). This notion should not be confused with the well-known cosmic censorship addressing naked-singularities. Our gedanken experiment applies far away from the spin and has nothing to do with singularities. Before we analyze different approaches to quantum gravity and discuss the existence or lack of a spin-censorship mechanism in each one, let us first show, just as a preface, why the gedanken experiment leads to a contradiction when analyzed with classical gravity.

### *Testing various classical and quantum gravity theories with the gedanken experiment*

In the rest of this article, we do not claim to find a single unquestionable mechanism that would provide a suitable spin-spacetime censorship, nor do we favor a single theory of classical or quantum gravity. Instead, we consider models of classical and quantum gravity proposed by various authors in the literature and test in each case whether it can model our gedanken experiment or whether it leads to a paradox. The first few censorship mechanisms we consider in the discussion (part 3) are contained within classical physics. These are meant to provide some intuition towards the presented gedanken experiment, but are far from being general and problem-free. Some of them modify the currently accepted theory of gravity. Specifically, they modify the EMFE or the stress-energy tensor, which of course have major implications on classical observables on cosmological scales, despite the small effect of the spin-induced curvature that led to these modifications. The rest of the censorship mechanisms in the discussion (part 4) involve different ways of incorporating quantum uncertainty into general relativity, such that the act of (quantum) measurement of spacetime prevents the spin from being determined, or causes back-action on the spin. Such theories are expected to provide the required censorship mechanism without altering the (classical) theory of gravity on cosmological scales. The latter approaches are therefore more plausible, but we nevertheless present briefly the former for the sake of completeness.

Eventually, the correct censorship mechanism must be derived from the yet unknown theory that governs the interaction between quantum spacetime and matter. **We show that by analyzing the requirements that the censorship mechanism must fulfill, it provides** **insights into how to properly describe quantum measurements of gravitational effects, which then provide new hints regarding the unified theory of quantum gravity.**

## 3. Discussion: classical approaches for spin-spacetime censorship

### *Classical approach #1 – Why classical gravity fails to describe the gedanken experiment, creating a paradox with relativistic causality*

The EMFE describe the coupling of the local spacetime curvature, expressed by Einstein's tensor $G^{\mu\nu}$ and the local stress-energy tensor $T^{\mu\nu}$, through the tensor equation $G^{\mu\nu} = \kappa T^{\mu\nu}$, where $\kappa = 8\pi G / c^4$ is Einstein's coupling constant. Applying these equations to a single electron in vacuum leads to an apparent paradox because our gedanken experiment requires $G^{\mu\nu}$ to be spherically symmetric, but the tensor $T^{\mu\nu}$ is aspherical due to the magnetic dipole moment of the electron. The axis of the magnetic dipole moment breaks the spherical symmetry of the Maxwell stress-energy tensor $T_{EM}^{\mu\nu} = \frac{-1}{\mu_0}\left(F^{\mu\alpha}F_{\alpha}^{\ \nu} + \frac{1}{4} g^{\mu\nu} F_{\alpha\beta} F^{\alpha\beta}\right)$ and therefore makes the combined tensor $T^{\mu\nu}$ aspherical. The resulting spacetime is known as the Kerr-Newman solution of the EMFE (see thorough discussion in[35,36]) but there are additional classical models[37] (also see SI section 7). The aspherical spacetime properties are apparent both in the far gravitational field (where we are not concerned with singularities) and in the near gravitational field of the electron (see Fig. 2). *As expected, the time dilation map is not spherically symmetric*, which seems to enable determining the particle's spin axis by measuring minuscule time dilation differences, leading to a paradox in our gedanken experiment. To avoid the apparent conflict between the EMFE and our gedanken experiment, there should exist some specific physical mechanism (most likely a quantum one) preventing the detection of deviations from spherical symmetry.

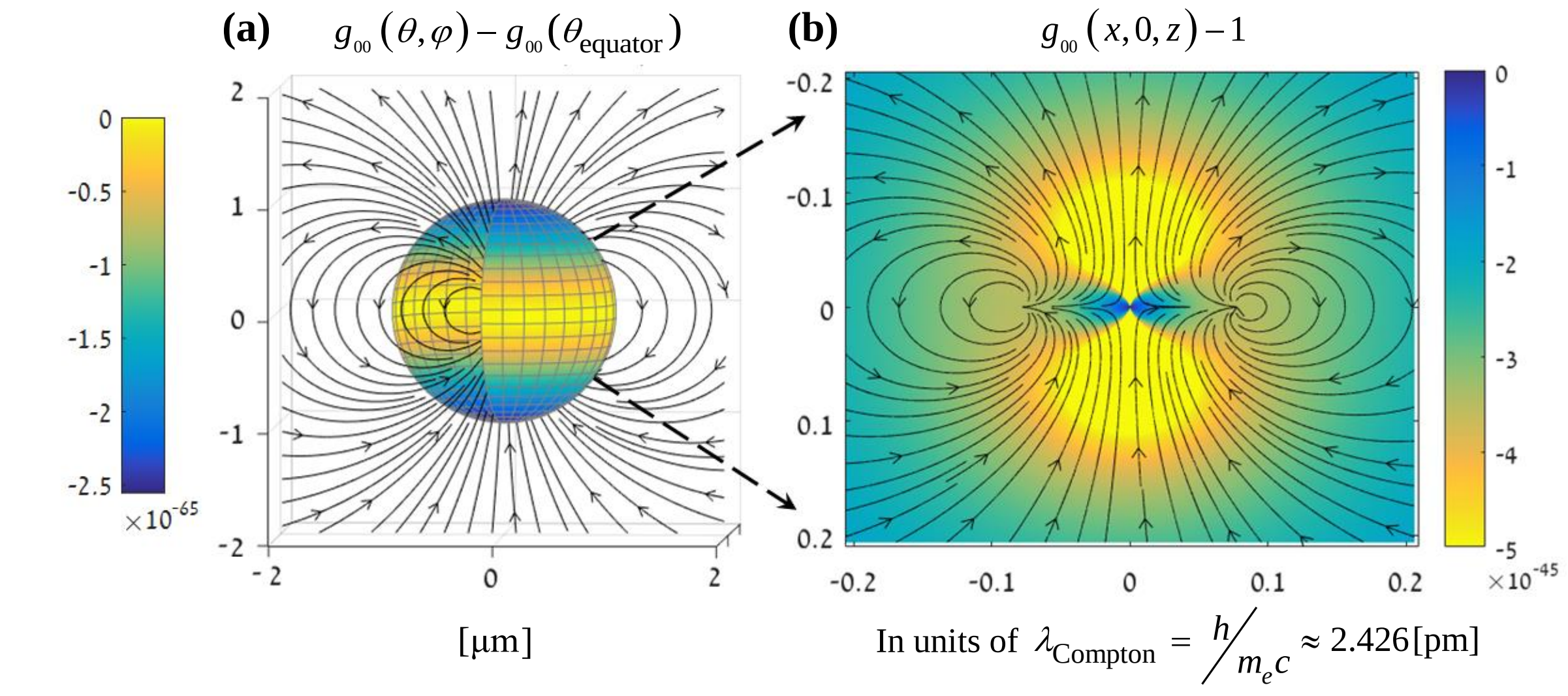

**Figure 2: The extent of time dilation around an electron according to the Einstein Maxwell field equations (EMFE)**, overlaid with its dipole magnetic field lines depicted in black: **(a)** The aspherical part of the time dilation distribution is depicted as a color map of $g_{00}(\theta,\varphi) - g_{00}(\theta_{\text{equator}})$ on a sphere of radius 1μm. Clocks at the equator are ticking faster than clocks at the north/south poles, and the time dilation is maximized when the magnetic dipole field is the strongest. **(b)** Zoomed-in ($\times 10^6$) plane cross-section of the time dilation (depicted as a color map of $g_{00}(x,0,z) - 1$). The time dilation is calculated according to the Kerr-Newman solution of the EMFE.

### *Classical approach #2 – Modifying the Maxwell stress-energy tensor*

Let us begin with a classical modification of general relativity that offers a resolution for maintaining relativistic causality (in the setup of our gedanken experiment). To obtain a spherically symmetric spacetime solution, we can replace the stress-energy tensor $T^{\mu\nu}$ in the EMFE with a hypothetical tensor $S^{\mu\nu}$ that is invariant under spatial rotations in the rest frame of a single isolated electron. This replacement ensures that the Einstein tensor, now obeying $G^{\mu\nu} = \kappa S^{\mu\nu}$, is also invariant under spatial rotations and therefore ensures a spherically symmetric spacetime, which resolves the apparent paradox. A possible choice for $S^{\mu\nu}$ is the stress-energy tensor associated with dust solutions[38], or null dust[39,40], i.e., $S^{\alpha\beta} = \sum_v \mathcal{E}(t,\mathbf{x};v^\mu) v^\alpha v^\beta$, where $\mathcal{E}(t,\mathbf{x};v^\mu)$ is the energy density of all the particles (including charged particles, photons and exchanged photons) passing at the point $(t,\mathbf{x})$ with velocity $v^\mu = dx^\mu / dt = (1, d\mathbf{x}/dt)$ (a quantum generalization should take into account the corresponding uncertainties). The dust stress-energy tensor includes all of the different particles' fluxes (including the photons that are exchanged between pairs of charged particles), to fulfill zero divergence $\nabla_\alpha S^{\alpha\beta} = 0$. For particles with non-zero rest mass, we have $\varepsilon v^\alpha v^\beta = \rho u^\alpha u^\beta$, where $\rho(t,\mathbf{x})$ is the proper density and $u^\alpha$ is defined as the four velocity (associated with $v^\alpha = dx^\alpha / dt$). Fig. 3 shows the resulting spacetime (for a single stationary particle), which is equivalent to the spherically symmetric Schwarzschild solution. In this way the paradox is avoided. Other helpful choices of tensor $S^{\mu\nu}$, some of which contain a natural spherical symmetry, can be proposed based on different stress-energy tensors (see e.g. the survey in[41]), and further generalizations, for instance to the Kaluza-Klein theory[42,43].

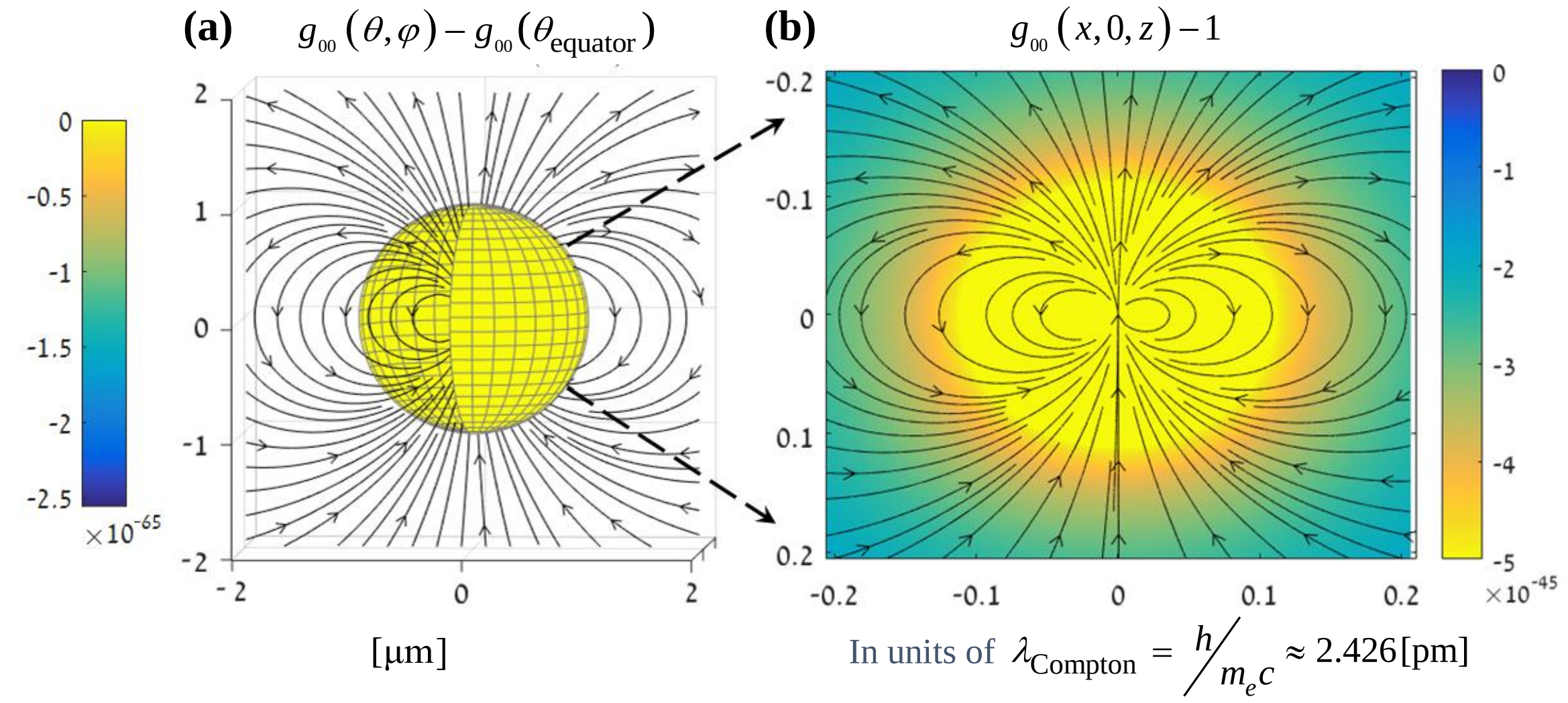

**Figure 3: The *measurable* time dilation should be spherically symmetric to prevent (or "censor"), precise assessment of the spin direction. (a)** A color map of $g_{00}(\theta,\varphi) - g_{00}(\theta_{\text{equator}})$ on a sphere of radius 1μm shows spherical symmetry, despite the magnetic dipole field lines overlaying the maps in black that are copied from Fig. 1. **(b)** Zoomed-in (×$10^6$) plane cross-section of the time dilation (depicted as a color map of $g_{00}(x,0,z) - 1$). Such a spherically symmetric spacetime may result from a modified EMFE or from a different stress-energy tensor (e.g., as in the dust stress-energy tensor approach). Here the time dilation is calculated according to the Schwarzschild metric, while the magnetic field lines are calculated in this curved spacetime, but do not influence it.

Replacing the stress-energy tensor $T^{\mu\nu}$ in EMFE with some hypothetical spherically symmetric tensor $S^{\mu\nu}$ carries significant consequences for classical physics; interestingly, it might soon be refuted in experiments: recent breakthroughs in measuring gravitational waves might bring soon the first observation of gravity waves induced by the *electromagnetic stress-energy tensor* (e.g., from newly born magnetars with extremely strong magnetic[44-48]).

### *Classical approach #3 – Adding a torsion tensor to EMFE*

It is important to ask whether previously proposed generalized forms of the Einstein equations may already contain some forms of censorship mechanisms. Many such variants of the Einstein equations have been discussed in the literature over the past century, e.g. the Einstein-Cartan theory[49,50], which creates a coupling between the intrinsic angular momentum (classical spin) of particles and the anti-symmetric part of the affine connection, known as the torsion tensor. Could the addition of torsion maintain the spherical symmetry by compensating for the particle spin? While the answer is in principle yes[51,52], it depends on the specific torsion tensor and it is unclear whether a single torsion tensor could compensate for any arbitrary particle spin. For example, it was proven that whenever the torsion is derived from a second-rank tensor potential, static spherically-symmetric solutions are not allowed[53], thus such torsion candidates would not suffice.

One may propose other candidate censorship mechanisms within classical physics. For example, an interesting (yet at this stage very speculative) idea is the complex electromagnetic tensor[54,55] that can eliminate the aspherical parts of the Maxwell stress-energy tensor. However, it remains to be seen whether such a theory is consistent with the electromagnetic theory. See SI section 5 for additional candidate classical approaches such as the possible existence of the electron electric dipole moment that fail to resolve the paradox.

## 4. Discussion: a quantum analysis of the gedanken experiment and some possible quantum approaches

One of the difficulties in formulating a quantum gravity theory is that quantum gravitational effects only appear at length scales near the Planck scale, around $10^{-35}$ m, a scale far smaller, and equivalently far larger in energy, than those currently accessible by high energy particle accelerators. Therefore, physicists lack experimental data, which could distinguish between competing theories and for this reason it is important to analyze gedanken experiments to see that they do not lead to contradictions.

All the above models used to analyze the gedanken experiment in a way that avoids a violation of causality also involve subtle alterations of the accepted EMFE (or the stress-energy tensor). Thus, they inevitably modify the accepted classical theory – but does it have to be so? Could there exist a quantum censorship mechanism that maintains a *measurable* spherical symmetry even when conventional classical gravity dictates asphericity? This question directly connects our gedanken experiment with the open questions regarding quantum measurement of spacetime in quantum gravity.

We now turn to discuss such censorship mechanisms that have to do with quantum corrections to the classical theory. Each of them reflects on some elements and ideas pertaining to the yet-to-be-found theory of quantum gravity. More generally, we analyze whether common elements from the literature on quantum measurement of spacetime could prevent the precise inference of the spin on the quantum level, and thus facilitate the missing censorship mechanism (preferably without altering EMFE in the classical limit). This way, we note which known approaches could facilitate the missing censorship mechanism and which seem inconsistent with our gedanken experiment.

We explore several types of non-classical approaches: The first approach treats spacetime as being classical, but the measurement device as being quantum, i.e., it regards the possible deviation from spherical symmetry *as a classical parameter which can be estimated using quantum metrology*[56]. The next three approaches treat spacetime as a fully quantum object that can be described by quantum operators, i.e., possible deviations from spherical symmetry are properties of the quantized spacetime that can become entangled with the spin. Here, the properties of the spacetime can be inferred using *a quantum measurement of the corresponding operator*, and they can cause back-action on the spin.

Furthermore, let us add several remarks about additional quantum theories that we have analyzed: We show that even using quantized gravity that takes into account the back-action on the spin (e.g., within the ADM formalism), our gedanken experiment still leads to a paradox with causality (see Methods section 4), proving that more advanced models are necessary. In addition, we analyze the perturbative approach of linearized quantum gravity, which may seem capable, in principle, of resolving causality paradoxes, but actually appears to lack the necessary mechanism in our case (detailed discussion in SI section 8).

### *Quantum approach #1 – Failure of quantum estimation due to decoherence*

A process of decoherence may limit the precision of estimating the spacetime parameters, thus possibly providing a censorship mechanism. For example, a depolarizing channel introduces isotropic loss of coherence, which makes the quantum state resemble a maximally mixed state[57]. This type of noise can therefore provide a suitable censorship mechanism, if applicable during our measurement of the quantum spacetime. Another example is a dephasing channel, which naturally arises when the system is immersed in an external fluctuating field[58] and can prevent the precise spin direction inference by limiting the measurement precision. Recently, a novel decoherence process was proposed, whose rate scales exponentially with the number of particles[59]. As will be shown below, such a strong decoherence process might be essential for masking the spin axis.

To quantify the precision limited by decoherence, we can treat the deviation from spherical symmetry, e.g., an axis-dependent time-dilation dictated by the metric element $g_{00}$, as a small parameter $\lambda$ that we wish to estimate using a quantum state $\rho_N$ of $N$ (possibly entangled) probing particles. Due to the measurement, the probe state undergoes some transformation $\Lambda_\lambda[\rho_N]$ yielding an estimate $\tilde{\lambda}$. For a general quantum state, there exists a bound (Cramér-Rao bound, see e.g.[60]) on the estimator's variance $V(\tilde{\lambda}) \geq 1/F_Q\left(\Lambda_\lambda[\rho_N]\right)$, that uses the quantum Fisher information (QFI) $F_Q$. We can now maximize the QFI over all states $\rho_N$ to find the limit on the quantum enhanced precision: $V(\tilde{\lambda}) \geq 1/F_N$, where $F_N = \max_{\rho_N} F_Q\left(\Lambda_\lambda[\rho_N]\right)$. A large number $N$ of probing particles can therefore estimate the spacetime around a single particle with increasing precision.

What can prevent the probes from finding the parameter $\lambda$ with a good enough precision that enables to infer the spin? The classical Fisher information scales like $N$, while the QFI[56] scales like $N^2$, enabling, in principle, better precision. However, it was shown[60] that in the presence of decoherence, this quantum enhancement diminishes. More generally, decoherence may grow with the number of probing particles and serve to limit the precision of the estimated $\tilde{\lambda}$. Therefore, we can speculate that the probing particles exert a non-negligible effect on the measured system and on the surrounding spacetime, thereby censoring the estimation of $\lambda$ or altering its value (making the scaling of the QFI much worse than $N^2$). A different measurement technique may involve a small number of probe particles but a prolonged probing time for increased precision in estimating the time-dilation difference. In this case, to prevent a precise estimation of $\lambda$ there could be a decoherence process which grows in time as quickly as the information about $\lambda$, thus bounding the estimation precision and providing the censorship mechanism[61].

## *Quantum approach #2 – Fluctuations of spacetime*

Another censorship mechanism relates the quantum uncertainty of time measurements with fluctuations of spacetime itself, upon the latter's quantization. An early version of this idea was suggested in the Quantum Foam model[62], which was later developed to loop quantum gravity[63,64] and spin foam[65]. These models employ quantum vacuum fluctuations that may prevent a precise time measurement from indicating a deviation from spherical symmetry. The challenge here is again to find the mechanism by which fluctuations consistently overcome the signal and hide deviation from spherical symmetry *even when we extend the duration of the time measurement*, or repeat it many times. Then we would be able to deduce that multiple couplings to spacetime also result in cumulative uncertainty that masks the spin value.

In light of the above, we can draw a general conclusion about the way quantum mechanics induces uncertainty into the gedanken experiment. To prevent the precise assessment of the spin axis, via repeated experiments (or one prolonged experiment), the uncertainty must grow as fast as the signal, an extraordinary behavior, as it seems to negate the Law of Large Numbers. If this behavior happens to be true in some scenarios resembling the aforementioned gedanken experiment, it may put a unique restriction on the sought-after theory of quantum gravity. It might be interesting to compare the above approach with stochastic quantum mechanics[66,67] advocating the inherent role of stochasticity in nature, conjectured to result from vacuum fluctuations[68].

## *Quantum approach #3 – Non-commutative geometry*

How else would it be possible to employ uncertainty in order to hide broken spherical symmetry? Another censorship mechanism could arise from having measurements of spacetime at different *locations* not commute with each other (analogous to measuring different components of the magnetic field). For example, if the time measurements by two clocks placed along the $\hat{\mathbf{x}}$ and $\hat{\mathbf{y}}$ axes are non-commutative, then it is *impossible* to measure time using these two clocks without uncertainty, thus preventing the determination of the spin axis. To quantify the amount of uncertainty it is worth considering a non-commutative Minkowski space that can be defined in terms of spacetime coordinates $x^\mu, \mu = 0,1,2,3$, which satisfy commutation relations of the form $[x^\mu, x^\nu] = i\Theta^{\mu\nu}$, where $\Theta^{\mu\nu}$ is some antisymmetric tensor. This non-commutativity implies the following uncertainty relations (see e.g.[69,70]) $\Delta x^\mu \Delta x^\nu \geq \frac{1}{2}\left|\Theta^{\mu\nu}\right|$. We thus see that, similar to measurements of a quantum electromagnetic field, one has to choose a measurement basis, thus rendering the system contextual. The chosen measurement basis breaks the spherical symmetry and prevents the measurement of the spin axis. (See the SI section 6 for further refinement arising from the case where the signaling protocol is performed with many pairs of entangled particles.)

## *Quantum approach #4 – Quantum decomposition into "plane waves"*

In this section, we present a quantum approach that may suggest, in its classical limit, EMFE with a modified stress-energy tensor (see also classical approach #1). To describe the coupling of an electron to the spacetime surrounding it, we consider first a ket state that represents a stationary (zero momentum – and thus completely delocalized) electron coupled to the spacetime degrees of freedom. This stationary delocalized electron ket state is considered to be a (spacelike uniform) superposition of infinitely many localized electron states. Each of these localized electron states is coupled to a quantum spacetime ket state with a specific metric corresponding to the position of the electron. We can then apply boost (affects the spacetime, as well as the electron's wavefunction) to obtain a quantum spacetime description of an electron with non-zero momentum. Finally, to construct a general quantum electron state, we use a superposition of these boosted states, creating an entangled state of the spin and the surrounding spacetime. This approach provides a candidate quantum description of an arbitrary electron state coupled to spacetimes. See SI section 3 for additional details. With this approach we can describe measurements of the spacetime by using a clock that measures the time dilation effects at a certain point. The clock and the electron are treated as parts of a single quantum mechanical system. Tracing out the electron degrees of freedom yields a density matrix that describes the clock. Using this density matrix, we can calculate the expectation value of the time dilation of the clock exerted by the electron's state (which is most generally described with a spinor wavefunction).

As before, measurements of time dilation via expectation values of clocks at different locations must not enable the paradoxical inference of the spin axis. Using this line of thought, we can ask ourselves – which requirement imposed on the quantum state will serve as the censorship mechanism here? Clearly, the expectation value of these measurement outcomes has to be spherically symmetric and independent of the spin axis. One possible resolution is similar to our #1 classical approach – modifying the Maxwell stress-energy tensor, forcing the spacetime engulfing a stationary electron to be spherically symmetric.

### *Other approaches that deserve further attention*

Many other candidate theories address the interface of gravity and quantum physics, and each of them could be used to model our gedanken experiment, then tested on whether it results in a causal paradox or not – which can now be used to test their validity. String theory must be considered, but there are other options as well. For example, linearized quantum gravity[71-73] may seem as a natural candidate, but a few difficulties seem to arise when trying to apply it to our gedanken experiment (see SI section 8). Similarly, in view of the back-action analysis presented in Methods section 4 it seems unclear how loop quantum gravity[63,64] could provide a suitable spin-spacetime censorship mechanism (perhaps there could be a different mechanism which is not based on back-action). As another example, gravitational decoherence – which in our problem means that the spacetime surrounding the electron spin "collapses" it – could affect our gedanken experiment via one of several quantitative models (e.g.,[74,75]). Recent advances in theories of gravitational decoherence offer intriguing thought and laboratory experiments that may resolve the conflict created by our gedanken experiment or be contradicted by it[76–79].

We should note the general problem with certain theories that involve nonlinear modifications of quantum mechanics – even the slightest ones lead to signaling[76–79]. However, recently it was shown how to avoid this problem in certain stochastic model[83] and thus, an effective nonlinear modification could potentially be based on these approaches. We restate the above findings in quantum gravity scenarios to study the implications of the quantum spin-spacetime coupling. Furthermore, we could speculate that using the stochasticity from the 2$^{nd}$ quantum approach above, we would arrive at re-linearized, non-signaling versions as well.

Finally, it could be interesting to examine our gedanken experiment in the context of the ER=EPR conjecture[84,85], according to which our EPR pair can be thought of as being connected by a Planckian wormhole.

## 5. Summary

As part of the ongoing search for a consistent interface between quantum mechanics and general relativity, we have analyzed a new gedanken experiment. The gedanken experiment is concerned with the way spacetime (classical or quantum) is altered by the spin-dependent electromagnetic fields that surround entangled electrons. The possibility of spacetime measurements that deviate from perfect spherical symmetry seems to violate relativistic causality and has led us to require a spin-spacetime censorship mechanism. Such a mechanism may lead to new restrictions on the way spacetime is distorted due to the presence of the stress-energy tensor (a more radical consequence of this work could be the inability to couple consistently spin with spacetime, even before introducing entanglement, quantum measurements, etc). Any mechanism proposed in the literature, speculative or mainstream, can now be tested for compatibility with our gedanken experiment, which could either challenge this mechanism or grant it further credence. **Therefore, the importance of our gedanken experiment is that it uses quantum information theory as a testing ground for existing and future theories of quantum measurement of spacetime and quantum gravity.**

## Methods

### *Method Section 1: Density matrix considerations*

This section analyzes the gedanken experiment in the language of density matrices, and discusses the underlying assumptions and their implications. Such a density-matrix-based description may be valid independently of the exact details of the unknown interaction Hamiltonian that couples the spin and spacetime degrees of freedom. Despite the generality of this description, it introduces nonlinearity at the level of the density matrix, which does not need not be the case with any model applied to our gedanken experiment.

First, let us write the density matrix in an EPR experiment: Alice uses a Stern-Gerlach device (oriented in a direction denoted by $\hat{\mathbf{n}}$) to measure the spin and there are two possible outcomes, $|+\hat{\mathbf{n}}\rangle$ or $|-\hat{\mathbf{n}}\rangle$. Thus, in the ordinary EPR experiment, it is well known that, regardless of Alice's choice, the reduced density matrix describing Bob's particle is: $\rho_{\text{Bob}} = \frac{1}{2}\left(|+\hat{\mathbf{n}}\rangle\langle+\hat{\mathbf{n}}| + |-\hat{\mathbf{n}}\rangle\langle-\hat{\mathbf{n}}|\right) = \frac{1}{2} I$, i.e., maximally mixed, for any choice of axis $\hat{\mathbf{n}}$ by Alice. However, if spacetime is treated in a quantum manner and if Bob uses clocks (instead of a Stern-Gerlach device), then there are additional quantum degrees of freedom, i.e., the spacetime changes due to the particle's spin.

At this point, we need to consider one detail of the interaction Hamiltonian that describes the mechanism of spin-spacetime coupling. For the rest of this section, we assume that the spacetime depends only on the spin axis, and not on its direction ($\pm$). This assumption may seem like an obvious choice at first glance, yet its consequences are significant. (In Methods section 4, for instance, we go beyond this assumption).

In order to measure the spin through the reading of the clocks' hands, Bob compares the delay $\tau_{|x|}$ of the clocks near the $\pm\hat{\mathbf{x}}$ axis with the time delay $\tau_{|y|}$ of the clocks near the $\pm\hat{\mathbf{y}}$ axis (see SI section 1 for discussion regarding parity symmetry of the spin's time dilation effect). To formalize the above in terms of density matrices, we assume in this section that the time dilation differences are correlated with Bob's spin axis (thus also correlated with Alice's choice of measurement) and schematically represented by the ket states $\left|\tau_{|x|}\right\rangle$ and $\left|\tau_{|y|}\right\rangle$. When Alice measures along the $\hat{\mathbf{x}}$ axis, the reduced density matrix becomes:

$$\rho^{x}{}_{\text{Bob}} = \tfrac{1}{2}\left(\left|\tau_{|x|}\right\rangle \otimes \left|+\hat{\mathbf{x}}\right\rangle\left\langle+\hat{\mathbf{x}}\right| \otimes \left\langle\tau_{|x|}\right| + \left|\tau_{|x|}\right\rangle \otimes \left|-\hat{\mathbf{x}}\right\rangle\left\langle-\hat{\mathbf{x}}\right| \otimes \left\langle\tau_{|x|}\right|\right) = \left|\tau_{|x|}\right\rangle\left\langle\tau_{|x|}\right| \otimes \tfrac{1}{2} I\,, \quad (1)$$

but if she measures along the $\hat{\mathbf{y}}$ axis, Bob's density matrix is different:

$$\rho^{y}{}_{\text{Bob}} = \tfrac{1}{2}\left(\left|\tau_{|y|}\right\rangle \otimes \left|+\hat{\mathbf{y}}\right\rangle\left\langle+\hat{\mathbf{y}}\right| \otimes \left\langle\tau_{|y|}\right| + \left|\tau_{|y|}\right\rangle \otimes \left|-\hat{\mathbf{y}}\right\rangle\left\langle-\hat{\mathbf{y}}\right| \otimes \left\langle\tau_{|y|}\right|\right) = \left|\tau_{|y|}\right\rangle\left\langle\tau_{|y|}\right| \otimes \tfrac{1}{2} I\,. \quad (2)$$

Importantly, the last two density matrices are no longer maximally mixed. Hence, if Bob has access to the spacetime degrees of freedom (e.g., via very precise clocks), he may use them instead of the spin degrees of freedom to decipher Alice's choice, thereby violating the no-signaling principle. Note that this violation is consistent with the literature[76–79].

We thus find that upon measurement, the spacetime state has to satisfy $\left|\tau_{|x|}\right\rangle = \left|\tau_{|y|}\right\rangle = \left|\tau\right\rangle$. It seems hard to avoid this requirement, which also arises when testing implications of quantum superpositions (SI section 3). Such a spherical symmetry is consistent, however, with classical theories that have a spherical spacetime, such as the dust stress-energy tensor (classical approach #2), which has to modify the classical EMFE. To find a consistent theory that does not force a spherical spacetime that is decoupled from the spin, we analyze the Dirac equation in curved spacetime and the ADM formalism (see Methods section 4), which however turn out to have their own limitations.

### *Method Section 2: Comparative analysis: inferring the spin from measuring its magnetic field – vs. – inferring the spin from measuring the curvature of spacetime*

It is well known that the axis and the direction of an unknown single spin-½ particle cannot be determined in general via measurements of its magnetic field. It seems to be a good idea to revisit the reasons why this cannot be done in order to understand what could be different in a spin measurement that is based on a time dilation (metrological) measurement. In this section, we show that there are inherent differences between the measurement of spacetime, e.g., with clocks, and the measurement of the magnetic dipole field, of a quantum spin. As with the clocks, the measurement of the spin's magnetic dipole field can be performed with measuring apparatuses coupled to the magnetic field and placed at different points in space.

When a single spin-½ state $|+\hat{\mathbf{s}}\rangle$ is measured through its magnetic interaction $\mathbf{B}\cdot\mu\mathbf{S}$, it is modified by the magnetic field $\mathbf{B}=B\hat{\mathbf{n}}$ of the measuring apparatus (the spin state $|+\hat{\mathbf{s}}\rangle$ evolves as a superposition of two nondegenerate states $|+\hat{\mathbf{n}}\rangle$ and $|-\hat{\mathbf{n}}\rangle$, with the corresponding eigenvalues of the self-adjoint operator $\mathbf{S}_{\hat{\mathbf{n}}}=\frac{1}{2}|+\hat{\mathbf{n}}\rangle\langle+\hat{\mathbf{n}}|-\frac{1}{2}|-\hat{\mathbf{n}}\rangle\langle-\hat{\mathbf{n}}|$). More generally, it can be shown that it is impossible to infer all of the components of a single spin-½ particle with magnetic field measurements because like the spin operators, $\mathrm{S}_x,\mathrm{S}_y,\mathrm{S}_z$, the magnetic field operators $\mathrm{B}_x,\mathrm{B}_y,\mathrm{B}_z$, do not commute. Recently, it was proposed how to estimate all three components of the magnetic field[86], but crucially, the fields were classical and quantum back-action (Methods section 4) was not applicable.

In contrast, quantitative commutation relations between operators describing the quantized *gravitational* field are currently unknown (and may not exist). As we do not know yet how to consistently and unambiguously quantize the gravitational field, we cannot rely on such uncertainty relations for providing a suitable spin-spacetime censorship mechanism. However, any future proposal for such a quantization method can be tested with our gedanken experiment to verify that it prevents the measurement of the spin axis.

Another interesting difference between the two types of measurements has to do with the sign $\pm$ of the spin direction. When clocks are used to measure the time-dilation effect, various theories yield results that are the same regardless of the sign of the spin direction $|+\hat{\mathbf{s}}\rangle$ or $|-\hat{\mathbf{s}}\rangle$ (i.e., the clocks show the same result even if the spin-½ is flipped). That is, in any such theory, **both of the orthogonal spin states $|+\hat{\mathbf{s}}\rangle$ and $|-\hat{\mathbf{s}}\rangle$ yield the same time dilation effect.** This degeneracy can be understood, for instance, in classical gravity theories, by the time-reversal symmetry properties of the spacetime metric $g_{\alpha\beta}(x^{\mu})$ (see SI section 1). The degeneracy of the time dilation measurement with respect to any given spanning ket basis $\{|+\hat{\mathbf{n}}\rangle, |-\hat{\mathbf{n}}\rangle\}$ means that this type of measurement does not act as a projection operator. However, once finding the axis without a projection, the entire spin state can be found without altering it, by placing a Stern-Gerlach magnets oriented along the spin axis (see SI section 10). Then, we get a contradiction with the "no-cloning" theorem. Therefore, we found a clear contrast with any measurement of the spin state through the magnetic field, which has to alter the spin state since it does not have the above degeneracy.

This degeneracy can be understood by investigating a clock's time dilation measurement as a metrological task, being an alternative to the conventional operator-based quantum mechanical spin state measurement. Attempts to describe the measurement of the spin via the time dilation of clocks seem to circumvent the limitation imposed by the commutation relation of quantum mechanical operators. Thus, the wavefunction of a spin-½ particle remains unchanged when it is measured with clocks in this metrological manner.

To summarize this section, the measurement of the spin through magnetic fields and through induced time dilations differ in a fundamental way. The former is known in quantum electrodynamics, while the latter depends on the yet unknown theory of quantum gravity. Despite not having the theory, certain general conclusion can be drawn, showing the necessity of a spin-½ spacetime censorship mechanism: For many candidate theories, the time dilation only depends on the spin axis and not on the spin direction. It follows that in all such theories, the spin axis can be fully determined with clocks, which leads to a paradox.

## *Method Section 3: Necessity of spin spacetime censorship for maintaining the principle of quantum superposition*

This section discusses the principle of quantum superposition in light of the spin-spacetime censorship. We model the gedanken experiment with a density matrix as in Methods section 1, and show how even relatively general quantum mechanical considerations still require the spacetime to be independent of the spin's axis (and the spin's direction). Below, we assume that the spin and its surrounding spacetime can be described separately (as a tensor product of states), and find the resulting conditions necessary for maintaining the principle of quantum superposition. To see this, consider a spin-½ charged fermion and describe the spin states of this fermion and its corresponding time dilation by: $\left|ST_{+x}\right\rangle = \left|+\hat{\mathbf{x}}\right\rangle \otimes \left|\tau_{|x|}\right\rangle$ and $\left|ST_{-x}\right\rangle = \left|-\hat{\mathbf{x}}\right\rangle \otimes \left|\tau_{|x|}\right\rangle$, with *ST* standing for the combined spin-spacetime quantum state, $\left|\pm\hat{\mathbf{x}}\right\rangle$ denoting the spin state (+½ or -½ with respect to the $\hat{\mathbf{x}}$ axis), and $\left|\tau_{|x|}\right\rangle$ denoting the time dilation effect associated with this spin axis. On the one hand, we know that a linear superposition of these two states should just be a spin pointing at the $+\hat{\mathbf{z}}$ axis direction – i.e., $\left|ST_{+z}\right\rangle = \left(\left|ST_{+x}\right\rangle + \left|ST_{-x}\right\rangle\right)/\sqrt{2}$, which is equal to $\left|ST_{+z}\right\rangle = \left|+\hat{\mathbf{z}}\right\rangle \otimes \left|\tau_{|z|}\right\rangle$. But on the other hand, $\left(\left|ST_{+x}\right\rangle + \left|ST_{-x}\right\rangle\right)/\sqrt{2} = \left(\left|+\hat{\mathbf{x}}\right\rangle \otimes \left|\tau_{|x|}\right\rangle + \left|-\hat{\mathbf{x}}\right\rangle \otimes \left|\tau_{|x|}\right\rangle\right)/\sqrt{2} = \left(\left|+\hat{\mathbf{x}}\right\rangle + \left|-\hat{\mathbf{x}}\right\rangle\right)/\sqrt{2} \otimes \left|\tau_{|x|}\right\rangle = \left|+\hat{\mathbf{z}}\right\rangle \otimes \left|\tau_{|x|}\right\rangle$. Consequently, $\left|\tau_{|x|}\right\rangle = \left|\tau_{|z|}\right\rangle$. More generally, using all possible linear superpositions of $\left|ST_{+x}\right\rangle = \left|+\hat{\mathbf{x}}\right\rangle \otimes \left|\tau_{|x|}\right\rangle$ and $\left|ST_{-x}\right\rangle = \left|-\hat{\mathbf{x}}\right\rangle \otimes \left|\tau_{|x|}\right\rangle$, it follows that the time-dilation effect must be spherically symmetric, and thus independent of the spin. It thus seems that the spin-½ algebra poses strong requirements on the descriptions of spacetime that *decouple* it from the spin, rendering it spherically symmetric around elementary spin-½ particles. It seems that in order to find a self-consistent theory that allows any spacetime-spin coupling, we have to consider a mechanism by which spacetime couples to the spin *direction* – this is analyzed in Methods section 4, together with the effect of the clock's back-action on the spin.

### *Method Section 4: Quantum back-action mechanisms that fail to provide a suitable spin-spacetime censorship mechanism*

Quantum gravitational back-action effect of the clocks on the spin-½ particle could in principle provide a suitable spin-spacetime censorship mechanism. In this section, we analyze these classical- and quantum-gravity back-action mechanisms:

I. A gravitational back-action effect, from the clocks' gravitational field, acting on the spin-½ Dirac spinor's gravitational dipole moment.

II. Analysis of back-action effects in quantized spacetime based on the definition of quantum operators for spacetime, as could be done with the ADM formalism, and assigning them with commutation relations that lead to the Wheeler-DeWitt equation.

Below we explain why these two back-action effects cannot provide a suitable spin-spacetime censorship mechanism.

**(I) Gravitational back-action acting on the spin-½ gravitational dipole moment.**

According to the Dirac equation in curved spacetime, spin-½ particles are expected to have a gravitational dipole moment[87]. This theory, although well established, is currently still hypothetical, because experiments that tested it showed inconclusive results (see e.g.,[88]). Furthermore, the theory is semi-classical (in the next subsection we quantize spacetime as well). Nevertheless, it is valuable to ask whether including the back-action effect of a spin-½ gravitational dipole moment could already provide the censorship mechanism, and thus model our gedanken experiment without contradicting causality. We will see below that this back-action mechanism fails to resolve the contradiction.

To explain how a back-action effect could be obtained, consider the clocks in our gedanken experiment and assume that these clocks, having spin 0, are ticking at a constant rate of $\omega_{\text{Clock}}$. The clocks slightly bend the (single, classical) spacetime around them as they must carry energy which is at least $\hbar\omega_{\text{Clock}}$. Now consider the back-action effect of the combined clocks' gravitational field $\vec{\mathbf{g}}_{\text{Clocks}}$ (in the rest frame of the spin-½ particle) and acting on the spin-½ gravitational dipole moment. Using the Dirac equation in curved spacetime (and neglecting high order relativistic terms in the electron's rest frame), the gravitational coupling effect on a spin-½ particle (with a spin pointing at direction $\hat{\mathbf{n}}$), is described by an interaction term of the form[87]):

$$H_{\text{int}} = \frac{\hbar}{2c}\vec{\mathbf{\Sigma}}\cdot\vec{\mathbf{g}}_{\text{Clocks}}, \tag{1}$$

with $\mathbf{\Sigma}_j = \begin{bmatrix} \boldsymbol{\sigma}_j & 0 \\ 0 & \boldsymbol{\sigma}_j \end{bmatrix}$ (where $\boldsymbol{\sigma}_1, \boldsymbol{\sigma}_2, \boldsymbol{\sigma}_3$ being the Pauli matrices).

This interaction changes the spin direction if $\hat{\mathbf{n}}$ and $\vec{\mathbf{g}}_{\text{Clocks}}$ are not collinear (and $\vec{\mathbf{g}}_{\text{Clocks}} \neq 0$). If the clock's time measurement (in our gedanken experiment) alters the spin direction through this back-action, our gedanken experiment could become analogous to the case of EPR, where Bob's measurement alters the spin and thus prevents the violation of causality. This interaction term thus seems to open a path to provide a self-consistent model to our gedanken experiment.

However, now consider the case in which the clocks are positioned symmetrically around the spin-½ particle so that the total gravitational field vanishes (i.e., $\vec{\mathbf{g}}_{\text{Clocks}} = 0$). Such a scheme could eliminate the back-action, and thus our gedanken experiment would still result in violation of causality. In particular, we analyze the case in which Bob's spin is prepared either along the X axis or along the Y axis. Bob uses symmetrically organized clocks such that the total gravitational field is zero, yet the spin axis still creates a different time dilation in the clocks depending on their positions (with respect to the spin axis - $\hat{\mathbf{n}}$ ). To be concrete, consider the case, in which Bob places six clocks symmetrically around his spin located at $\vec{\mathbf{r}}_{\text{Bob's Spin}}$ , i.e., the clocks are at positions $\vec{\mathbf{r}}_{\text{Bob's Spin}} \pm L\hat{\mathbf{x}}$, $\vec{\mathbf{r}}_{\text{Bob's Spin}} \pm L\hat{\mathbf{y}}$ and $\vec{\mathbf{r}}_{\text{Bob's Spin}} \pm L\hat{\mathbf{z}}$ (which should be understood as position expectation values since the spin and clocks are still non-classical). At short time scales, the symmetric positioning would result in a zero gravitational field on the spin. What about longer time scales?

While initially it seems that there is no back-action, we note that due to the gravitational pull (of the spin's gravitation dipole moment) the clocks are expected to change their positions. In particular, the two clocks which are aligned along the spin axis are expected to be pulled in an asymmetric manner (one of them will get closer to the spin, while the other will get farther from it). Thus, the total gravitational field on the spin becomes nonzero.

Furthermore, the other four clocks, on the axes orthogonal to the spin, could potentially be influenced by a frame dragging effect of the spin's gravitational dipole moment[89]. Importantly, due to symmetry considerations, these four clocks maintain a symmetric configuration around the spin and the whole configuration (six clocks and spin) remain symmetric with respect to a 90 degrees rotation around the spin axis. Thus, the total gravitational field of the clocks, $\vec{\mathbf{g}}_{\text{Clocks}}$ remains parallel to the spin axis. That is, if the spin is prepared in the $\pm\hat{\mathbf{x}}$ direction, the clocks positions evolve so that the total gravitational field is $\vec{\mathbf{g}}_{\text{Clocks}} = \pm g\hat{\mathbf{x}}$, and if the spin is prepared in the $\pm\hat{\mathbf{y}}$ direction, the clocks positions evolve so that the total gravitational field is $\vec{\mathbf{g}}_{\text{Clocks}} = \pm g\hat{\mathbf{y}}$. Most importantly, due to symmetry considerations, and even after the evolution of the clocks' wavefunctions, the spin direction $\hat{\mathbf{n}}$ and the gravitational field $\vec{\mathbf{g}}_{\text{Clocks}}$ remain aligned. This alignment can also be understood in another way: with angular momentum considerations. **Either way, we conclude that the spin axis remains unchanged**. Indeed, at all times, the spin state $\left|\hat{\mathbf{n}}\right\rangle$ (where $\hat{\mathbf{n}}=\pm\hat{\mathbf{x}}$ or $\hat{\mathbf{n}}=\pm\hat{\mathbf{y}}$) is an eigenstate of the interaction term $\frac{\hbar}{2c}\vec{\boldsymbol{\Sigma}}\cdot\vec{\mathbf{g}}_{\text{Clocks}}$, since, $\frac{\hbar}{2c}\vec{\boldsymbol{\Sigma}}\cdot\vec{\mathbf{g}}_{\text{Clocks}}\left|\hat{\mathbf{n}}\right\rangle=\frac{\hbar}{2c}g_{\text{Clocks}}\left(\vec{\boldsymbol{\Sigma}}\cdot\hat{\mathbf{n}}\right)\left|\hat{\mathbf{n}}\right\rangle= \;=\frac{\hbar}{2c}g_{\text{Clocks}}\left|\hat{\mathbf{n}}\right\rangle$.

We can summarize this subsection with the conclusion that we found a scenario in which the back-action on the spin does not alter the axis of the spin due to symmetry arguments. Therefore, it seems like Bob is able to determine the spin axis ($\pm\hat{\mathbf{x}}$ axis, or $\pm\hat{\mathbf{y}}$ axis) without altering this axis – which leads to a contradiction with relativistic causality. We conclude from the above discussion that the Dirac equation in curved spacetime and the theory of gravitational dipole moment do not provide a suitable spin-spacetime censorship mechanism.

Note that the same arguments above actually **lead to the same conclusion in a much more general framework**: It does not matter if the spin has a gravito-dipole-moment, or only a quadropole moment, or another higher order multipole. The logic that led to the paradox works for *any* response to the gravitational field. Regardless of the mechanism, having a single classical spacetime that interacts with all the clocks can be used to place several clocks that cancel out the field on the spin. Thus, the above conclusion is valid beyond the Dirac equation in curved spacetime. In the next subsection we find that similar symmetry arguments lead to similar conclusions even for fully quantized theories of gravity.

**(II) Gravitational back-action by quantized spacetime**

In this subsection, we shall generalize the discussion of the previous subsection, to the case of a fully quantized spacetime. For this purpose we shall revisit the case in which Bob places six clocks symmetrically around his spin located at $\vec{\mathbf{r}}_{\text{Bob's Spin}}$, i.e., the clocks are described by wave-packets located symmetrically at positions $\vec{\mathbf{r}}_{\text{Bob's Spin}}\pm L\hat{\mathbf{x}}$, $\vec{\mathbf{r}}_{\text{Bob's Spin}}\pm L\hat{\mathbf{y}}$ and $\vec{\mathbf{r}}_{\text{Bob's Spin}}\pm L\hat{\mathbf{z}}$ (and of course remember that Bob's spin was prepared either along the $\hat{\mathbf{x}}$-axis or along the $\hat{\mathbf{y}}$-axis). Now, let us analyze the interaction of the clocks with the spin through a general framework of a quantized spacetime.

It turns out, that there is a basic symmetry argument based on angular momentum conservation, which leads to the very general conclusion: The quantization of spacetime cannot provide a back-action mechanism that can alter the spin axis provided the clocks are arranged symmetrically around it as described above. To explain the argument, consider the total angular momentum of the six spin-0 clocks with the spin-½ particle. Initially, the orbital angular momentum of the six clocks is $\vec{\mathbf{L}}_{orbital} = 0$ and the spin's angular momentum is $\vec{\mathbf{S}}_{spin} = ½\hbar\hat{\mathbf{n}}$ for the spin-½ particle. Therefore, the total angular momentum of the system is $\vec{\mathbf{J}}_{\mathrm{Tot}}(t=0) = ½\hbar\hat{\mathbf{n}}$. After a while, due to possible gravitational back-action effects, the spin part of the angular momentum and the orbital part of the angular momentum exchange angular momenta, such that they can become entangled. However, the total angular momentum is preserved. The state of the system is generally,

$$\left|\vec{\mathbf{J}}_{\mathrm{Tot}}(t)\right\rangle = \alpha_{+}(t)\left|\vec{\mathbf{S}}_{spin} = ½\hbar\hat{\mathbf{n}}\right\rangle \otimes \left|\vec{\mathbf{L}}_{orbital} = 0\right\rangle + \alpha_{-}(t)\left|\vec{\mathbf{S}}_{spin} = -½\hbar\hat{\mathbf{n}}\right\rangle \otimes \left|\vec{\mathbf{L}}_{orbital} = \hbar\hat{\mathbf{n}}\right\rangle, \qquad (2)$$

with $|\alpha_{+}(t)|^{2} + |\alpha_{-}(t)|^{2} = 1$.

We can no longer describe the spin state as being pure, pointing along a certain direction, since it is entangled to the quantum state of the clocks. Generally, the relative values of $\alpha_{+}(t), \alpha_{-}(t)$ should describe the spin rotation and dynamics in time (if not for its entanglement with the clocks). Nevertheless, there is no mechanism breaking the symmetry, and thus, the spin remains along the same axis. Even if the spin flips (due to gravitational back-action spin-orbital angular momenta exchange), the spin axis remains aligned with its original orientation (i.e., along the $\hat{\mathbf{n}} = \pm\hat{\mathbf{x}}$-axis or along the $\hat{\mathbf{n}} = \pm\hat{\mathbf{y}}$-axis). Furthermore, the entanglement of the spin axis with the orbital angular momentum, and through it with any additional degrees of freedom – e.g., clocks positions, quantized spacetime degrees of freedom, etc., prevents its superposition, so it stays aligned along its original axis. We conclude that this back-action mechanism cannot provide the needed spin-spacetime censorship that would avoid the violation of causality: Bob can find out the axis of the spin by measuring the difference in time dilations between the clocks, which themselves only alter the spin direction and not its axis.

Let us now demonstrate these ideas using the ADM formalism. After applying the ADM quantization, we shall arrive to the Wheeler-DeWitt equation, which describes both matter and spacetime using one wavefunction $\Psi\left[h_{ij},\varphi_{\text{Matter}}\right]$, where $h_{ij}$ corresponds to the metric tensor part of the spatial foliation slices of a spacetime. i.e., $g_{\mu\nu}dx^{\mu}dx^{\nu} = = (-N^2+\beta_k\beta^k)dt^2 + 2\beta_k dx^k dt + h_{ij}dx^i dx^j$ and $\varphi_{\text{Matter}}$ denotes the mater state, i.e., $\varphi_{\text{Matter}}$ includes the electron's spinor and clocks' scalar fields.

We apply the ADM formalism to the Einstein-Hilbert action $S = \int\left[\frac{1}{2\kappa}R + \mathcal{L}_{\text{M}}\right]\sqrt{-g}\,\mathrm{d}^4x$, where $R$ is the Ricci scalar, $\mathcal{L}_{\text{M}}$ is a term describing particle fields, and $\kappa = 8\pi G/c^4$ is the Einstein's coupling constant. The result is a Hamiltonian of the form $H_{\perp} = \frac{1}{2\sqrt{h}}G_{ijkl}\pi^{ij}\pi^{kl} - \sqrt{h}\,^{(3)}R$, where $h = \det(h_{ij})$, $G_{ijkl} = (h_{ik}h_{jl} + h_{il}h_{jk} - h_{ij}h_{kl})$ is the Wheeler–DeWitt metric, $^{(3)}R$ is the induced curvature (within the spatial foliation slice), and $\pi^{ij}$ are the conjugate momenta. The quantization of the ADM Hamiltonian means that the conjugate momenta are interpreted as operators satisfying $\pi^{ij}(t,\mathbf{x}) := -i\frac{\delta}{\delta h_{ij}(t,\mathbf{x})}$, and having the usual quantum commutation relations with the $h_{ij}(t,\mathbf{x})$. Finally, the Wheeler DeWitt equation is

$$H_{\perp}\Psi\left[h_{ab},\varphi_{\text{Matter}}\right] = \left(\frac{1}{2\sqrt{h}}G_{ijkl}\frac{\delta}{\delta h_{ij}(t,\mathbf{x})}\frac{\delta}{\delta h_{kl}(t,\mathbf{x})} - \sqrt{h}\left(\frac{1}{2\kappa}\,^{(3)}R + \mathcal{L}_{\text{M}}\right)\right)\Psi\left[h_{ab},\varphi_{\text{Matter}}\right] = 0. \quad (3)$$

We can now use this equation to re-examine our arguments regarding the idea of back-action mechanism from gravitational dipole moment (associated with Dirac equation in curved spacetime). To do that, we consider the total 4-spinor wavefunction $\varphi_{\text{Matter}} = \psi^{\mu}(t, \mathbf{x}_s, \mathbf{x}_1, \mathbf{x}_2, \mathbf{x}_3, ..., \mathbf{x}_6)$ describing the spin-½ and the six identical spin 0 clocks around it. We note that $\psi^{\mu}(t, \mathbf{x}_s, \mathbf{x}_1, \mathbf{x}_2, \mathbf{x}_3, ..., \mathbf{x}_6) = \psi^{\mu}(t, \mathbf{x}_s, \mathbf{x}_{p(1)}, \mathbf{x}_{p(2)}, \mathbf{x}_{p(3)}, ..., \mathbf{x}_{p(6)})$ for every permutation $j \to p(j)$ of the clocks' coordinates, because the clocks are identical bosons. It is of course very difficult to solve this equation, but it is enough to verify the angular momentum argument on it. Due to conservation of angular momentum, if the spin is pointing at direction $+\hat{\mathbf{n}}$ and the spin 0 clocks around it are initially stationary, the total angular momentum of the system is constant at all subsequent times and equal to $\mathbf{J} = +½\hbar\hat{\mathbf{n}}.$ The next paragraph shows how conservation of angular momentum forbids the spin axis change by the back-action from the clocks.

Consider now the solution of Wheeler-DeWitt equation. In particular, consider the case that the six clocks are described by wave-packets located at positions $\pm L\hat{\mathbf{x}}$, $\pm L\hat{\mathbf{y}}$ and $\pm L\hat{\mathbf{z}}$, with respect to the position of the spin (with the spin prepared at $\hat{\mathbf{n}} = \pm\hat{\mathbf{x}}$ or $\hat{\mathbf{n}} = \pm\hat{\mathbf{y}}$). Importantly, the combined state of the clocks and the spin-½ is an eigenstate of a 90° rotation around the $+\hat{\mathbf{n}}$ axis (i.e., $R_{-90^o}(\hat{\mathbf{n}})\psi^{\mu}(t_0, \mathbf{x}_s, \mathbf{x}_1, \mathbf{x}_2, \mathbf{x}_3, ..., \mathbf{x}_6) = \mathrm{e}^{-i\pi/4}\psi^{\mu}(t_0, \mathbf{x}_s, \mathbf{x}_1, \mathbf{x}_2, \mathbf{x}_3, ..., \mathbf{x}_6)$, where the phase $\mathrm{e}^{-i\pi/4}$ is due to the half integer angular momentum $\mathbf{J} = +½\hbar\hat{\mathbf{n}}.$ It follows that a $\Psi\left[h_{ab}, \varphi_{\text{Matter}}\right]$ solution of (3) is also an eigenstate of the 90° rotation around the $+\hat{\mathbf{n}}$ axis. Furthermore, with proper preparation, the initial state of $\Psi\left[h_{ab}, \varphi_{\text{Matter}}\right]$ is an eigenstate of the unitary $R_{-90^o}(\hat{\mathbf{n}})$ rotation operator. Thus, according to (3), for any interaction of the clocks and the spin, with any induced entanglement of the spacetimes' spatial metrics $h_{ab}$ with $\varphi_{\text{Matter}}$, the state remains an eigenstate of the unitary $R_{-90^o}(\hat{\mathbf{n}})$. This simple symmetry argument proves that if the spin is located at direction $\hat{\mathbf{n}}$, it will remain along this direction, only able to flip, but not to rotate. Therefore, Bob's measurement does not alter the spin axis.

How does Bob measure the spin axis? The time dilation measured values of the four clocks in positions perpendicular to $\hat{\mathbf{n}}$ must all be the same. Thus, by comparing the time dilations of the clocks in $\pm\hat{\mathbf{x}}$ and in $\pm\hat{\mathbf{y}}$ with the clocks in $\pm\hat{\mathbf{z}}$, Bob can find out the axis of the spin. Since this measurement is done without altering the axis, we conclude that the ADM formalism does not provide a suitable back-action censorship mechanism.

# Supplementary Information

**Table of Contents**

## *Section 1: Symmetry properties of the spacetime metric and measurements of all its components*

In this section we discuss whether the spin value (up or down), rather than its axis only, has any effect on spacetime. Up to now we have mostly considered $g_{00}(x^{\mu})$, which is often the most significant component of the metric, but to answer this question we shall now study all the 16 $g_{\alpha\beta}(x^{\mu})$ terms, and analyze their symmetry properties.

Let us first discuss parity and time-reversal transformations. The parity transformation defined by $P(x^0, x^1, x^2, x^3) \to (x^0, -x^1, -x^2, -x^3)$ and the time-reversal transformation defined by $T(x^0, x^1, x^2, x^3) \to (-x^0, x^1, x^2, x^3)$ flip the sign of the spin. However, due to the tensor properties of the spacetime metric $g_{\alpha\beta}(x^{\mu})$, both $g_{00}$ and all of the $g_{ij}$ $(1 \le i, j \le 3)$ remain invariant under both parity and time-reversal transformations, i.e., $g_{00}(x^{\mu}) = g_{00}(P(x^{\mu})) = g_{00}(T(x^{\mu}))$ and $g_{ij}(x^{\mu}) = g_{ij}(P(x^{\mu})) = g_{ij}(T(x^{\mu}))$. Since the spin flips its sign under these transformations, it follows that $g_{00}$ and $g_{ij}$ $(1 \leq i, j \leq 3)$ are independent with respect to the spin sign.

In contrast, the six $g_{i0}, g_{0i}$ components flip their signs under parity and time-reversal transformations:

$g_{i0}(x^{\mu}) = -g_{i0}(P(x^{\mu})) = -g_{i0}(T(x^{\mu})) = g_{0i}(x^{\mu}) = -g_{0i}(P(x^{\mu})) = -g_{0i}(T(x^{\mu}))$. Since the spin also flips its sign under these transformations, the six $g_{i0}, g_{0i}$ components are either (#1) correlated to the spin or (#2) nonexistent, i.e., equal to zero (if, in addition, $g_{00}, g_{ij}$ are time-independent, then the spacetime would be called static). If $g_{i0}, g_{0i}$ are nonzero and correlated to the spin, as in option (#1), then there should exist a stronger spin censorship mechanism preventing not just the detection of the spin axis (as in the main text) but also the detection of the spin direction. In the next paragraphs we prove that in order to prevent a paradox, the measurable spacetime and all components of $g_{\alpha\beta}(x^{\mu})$ are both spherically symmetric and static, with $g_{i0}, g_{0i}$ being zero, i.e., option (#2).

To continue, we show that all 16 terms of $g_{\alpha\beta}(x^{\mu})$ can be measured with clocks, each moving along a different timelike trajectory in spacetime. Denote the trajectory of the clock as $x^{\mu}(t)$. The proper-time interval $d\tau$ along a short timelike 4-vector interval $dx^{\mu}$ of the trajectory $x^{\mu}(t)$ is given by $c^2 d\tau^2 = g_{\alpha\beta}(x^{\mu}(t)) dx^{\alpha} dx^{\beta}$. Thus, by using several variations of the clock's trajectory, one can determine *all* of the components of the spacetime metric $g_{\alpha\beta}(x^{\mu}(t))$. This is true, because $g_{\alpha\beta} = g_{\beta\alpha}$ and also because tensors of the form $dy^{\alpha} dy^{\beta}$ (where $dy^{\alpha}$ are time-like 4-vectors) span the space of (4×4) symmetric matrices (due to symmetry properties of the spacetime, one can greatly reduce the number of necessary measurements).

The ability to use clocks for measuring all the components of the spacetime metric $g_{\alpha\beta}(x^{\mu}(t))$ suggests a generalized gedanken experiment that follows the same lines of the experiment in the main text, and implies that there exists a generalized spin spacetime censorship: It should be impossible to infer the spin axis and direction (of a spin-½ particle) from measuring any component of $g_{\alpha\beta}(x^{\mu}(t))$ and from any measurement of proper-time intervals $c^2 d\tau^2 = g_{\alpha\beta}(x^{\mu}(t))dx^{\alpha}dx^{\beta}$. Interestingly, such a generalized spin spacetime censorship principle imposes additional important constraints on the measurable values of the spacetime metric $g_{\alpha\beta}$ around a spin-½ particle. The measurable tensor has to be spherically symmetric, and static with the six components $g_{i0}, g_{0i}$ being zero. These constraints and other additional symmetry requirements will be presented below.

We introduce two additional symmetry properties of the spacetime metric that are associated with the spin: (1) invariance of the metric tensor with respect to continuous time translations; (2) invariance of the metric tensor with respect to continuous rotations around the spin axis. A spacetime endowed with these two symmetries is both stationary and cylindrically symmetric. With a proper choice of spherical coordinate system $(t, r, \theta, \phi)$, it is well known that the most general possible form of this metric is given by[90],

$$c^2 d\tau^2 = e^{2\lambda(r,\theta)} c^2 dt^2 - e^{2\eta(r,\theta)} dr^2 - e^{2\psi(r,\theta)} r^2 d\theta^2 - e^{2\tilde{\psi}(r,\theta)} r^2 \sin(\theta)^2 \left(d\phi - \omega(r,\theta)dt\right)^2, \quad (4)$$

where $\theta$ is the polar angle measured with respect to the spin axis and $\phi$ is the azimuthal angle (describing rotations around the axis of the spin). We can then measure the components of this spacetime metric, with clocks moving along a specific timelike 4-vector in the spacetime.

Choosing for the clocks stationary 4-vector paths of the form $(t_0, r_0, \theta_0, \phi_0) + (dt, 0, 0, 0)$, we see that spin spacetime censorship is maintained if and only if $\lambda(r,\theta) = \lambda(r)$ - i.e., $\lambda$ is only a function of $r$ (independent of $\theta$). Furthermore, choosing radial paths for the clocks, $(t_0, r_0, \theta_0, \phi_0) + (dt, u_r dr, 0, 0)$ with $|u_r| < c$, we see that spin spacetime censorship is maintained if and only if $\eta(r,\theta) = \eta(r)$. Finally, choosing spherical tangent trajectories of the form $(t_0, r_0, \theta_0, \phi_0) + (dt, 0, u_\theta dt, u_\phi dt)$ with $r_0^2 u_\theta^2 + r_0^2 \sin(\theta_0)^2 u_\phi^2 = u_{\text{clock}}^2 < c^2$ we see that in order to maintain the spin spacetime censorship, the proper-time $d\tau$ must be a function of just two parameters: the radial coordinate $r$ and the speed $u_{\text{clock}}$. It follows that we must have $\tilde{\psi}(r,\theta) = \psi(r,\theta) = \psi(r)$ and $\omega(r,\theta) = 0$. Thus, spin-spacetime censorship seems to allow only spacetime metrics equivalent, upon measurement, to $c^2 d\tau^2 = e^{2\lambda(r)} c^2 dt^2 - e^{2\eta(r)} dr^2 - e^{2\psi(r)} \left( r^2 d\theta^2 + r^2 \sin(\theta)^2 d\phi^2 ) \right)$, which implies spherical symmetry (for a precise definition see e.g., [91,92]). This spherical symmetry of the spacetime guarantees that the spin's axis and the spin's direction cannot be inferred with clocks regardless of their trajectories.

To summarize, this section showed that generalizing our gedanken experiment enables using a clock to measure all the components of $g_{\alpha\beta}(x^\mu)$. In principle, the terms $g_{i0}, g_{0i}$ should enable measuring the direction of the spin, and not only its axis, as in the main text. Therefore, a classical way to prevent a paradox in our gedanken experiment is by determining that the measurable spacetime and all the $g_{\alpha\beta}(x^\mu)$ components are spherically symmetric and static, with $g_{i0}, g_{0i}$ being zero. As explained in the main text, some quantum approaches could potentially bypass a small classical deviation from spherical symmetry.

*Section 2: A related gedanken experiment with spatial degrees of freedom instead of spin*

This section, which continues the theme of combining concepts from general relativity and quantum information, describes a related gedanken experiment. In this additional gedanken experiment, the entangled variables of Alice and Bob are their particles' positions instead of spins. This gedanken experiment is a modified version of the one presented in[93]. It could be utilized to present some well-known conceptual problems with the semi-classical model (according to which it is assumed that the spacetime curvature is proportional to the expectation value of the stress-energy tensor[93,94]). While the semi-classical approximation creates a paradox with the gedanken experiment presented in this section, we will show that a simple quantization of linearized gravity leads to physically sound results (no paradox), namely that relativistic causality is maintained. We also discuss the difference between the two experiments, i.e. why the original gedanken experiment from the main text is not resolved (i.e., causality and no cloning are not maintained) by such a quantization of linearized gravity (in which each of the particles' ket states couples to a different spacetime ket state).

To analyze this gedanken experiment, consider an entangled state of the form

$$|\psi\rangle = \frac{1}{\sqrt{2}}\left(\left|\mathbf{o}_B + \mathbf{R_p}\right\rangle_B \otimes \left|\mathbf{o}_A + \mathbf{R_p}\right\rangle_A + \left|\mathbf{o}_B - \mathbf{R_p}\right\rangle_B \otimes \left|\mathbf{o}_A - \mathbf{R_p}\right\rangle_A\right) =$$
$$\equiv \frac{1}{\sqrt{2}}\left(|+\mathbf{B}\rangle_B \otimes |+\mathbf{A}\rangle_A + |-\mathbf{B}\rangle_B \otimes |-\mathbf{A}\rangle_A\right), \tag{5}$$

where $\left|\mathbf{o}_A + \mathbf{R_p}\right\rangle_A \equiv |+\mathbf{A}\rangle_A$ and $\left|\mathbf{o}_A - \mathbf{R_p}\right\rangle_A \equiv |-\mathbf{A}\rangle_A$ denote two discrete ket state particle positions in Alice's lab (and similarly, $|\mathbf{o}_B + \mathbf{B}\rangle_B \equiv |+\mathbf{B}\rangle_B$, $|\mathbf{o}_B - \mathbf{B}\rangle_B \equiv |-\mathbf{B}\rangle_B$ denote discrete ket state particle positions in Bob's lab).

In an attempt to communicate with Bob, Alice can measure her particle using any basis she wants to choose. Then, Bob can use clocks to measure time-dilation effects in an attempt to decipher Alice's choice of measurement basis. Using the linear approximation of gravity we find that, $g_{00,+\mathbf{B}}(\mathbf{r}) \approx \eta_{00} + h_{00}(\mathbf{r}) = 1 + \frac{2G}{c^4}\left( mc^2 \frac{1}{|\mathbf{r}-\mathbf{B}|} \right)$ if Bob's particle is in $|+\mathbf{B}\rangle_B$ and $g_{00,-\mathbf{B}}(\mathbf{r}) \approx \eta_{00} + h_{00}(\mathbf{r}) = 1 + \frac{2G}{c^4}\left( mc^2 \frac{1}{|\mathbf{r}+\mathbf{B}|} \right)$ if his particle is in $|-\mathbf{B}\rangle_B$. Thus, the combined state of the system (taking into account Bob's time dilation measurements) is described by

$$|\psi\rangle = \frac{1}{\sqrt{2}}\left( |g_{00,+\mathbf{B}}(\mathbf{r})\rangle \otimes |+\mathbf{B}\rangle_B \otimes |+\mathbf{A}\rangle_A + |g_{00,-\mathbf{B}}(\mathbf{r})\rangle \otimes |-\mathbf{B}\rangle_B \otimes |-\mathbf{A}\rangle_A \right). \tag{6}$$

Now let us consider the effect of Alice's choice of measurement basis. Alice can choose any basis of the form $\left\{ \alpha|+\mathbf{A}\rangle_A + \beta|-\mathbf{A}\rangle_A,\ \beta^*|+\mathbf{A}\rangle_A - \alpha^*|-\mathbf{A}\rangle_A \right\}$. The effect of her measurement on Bob's particle is described by tracing out the density matrix $\rho = |\psi\rangle\langle\psi|$ with respect to her measurement basis. Noting that

$$\left( \alpha^* \langle +\mathbf{A}|_A + \beta^* \langle -\mathbf{A}|_A \right)|\psi\rangle = \frac{1}{\sqrt{2}}\left( |g_{00,+\mathbf{B}}(\mathbf{r})\rangle \otimes \alpha^* |+\mathbf{B}\rangle_B + |g_{00,-\mathbf{B}}(\mathbf{r})\rangle \otimes \beta^* |-\mathbf{B}\rangle_B \right)$$

$$\left( \beta \langle +\mathbf{A}|_A - \alpha \langle -\mathbf{A}|_A \right)|\psi\rangle = \frac{1}{\sqrt{2}}\left( |g_{00,+\mathbf{B}}(\mathbf{r})\rangle \otimes \beta |+\mathbf{B}\rangle_B - |g_{00,-\mathbf{B}}(\mathbf{r})\rangle \otimes \alpha |-\mathbf{B}\rangle_B \right) \tag{7}$$

we obtain a reduced density matrix of the form

$$\begin{aligned} \rho_B &= \mathrm{Tr}_A \rho = {}_A\langle +\mathbf{A}|\rho|+\mathbf{A}\rangle_A + {}_A\langle -\mathbf{A}|\rho|-\mathbf{A}\rangle_A = \\ &= \frac{1}{2}\Big[ \left( |g_{00,+\mathbf{B}}(\mathbf{r})\rangle \otimes \alpha^* |+\mathbf{B}\rangle_B + |g_{00,-\mathbf{B}}(\mathbf{r})\rangle \otimes \beta^* |-\mathbf{B}\rangle_B \right)\left( \langle g_{00,+\mathbf{B}}(\mathbf{r})| \otimes \alpha\ {}_B\langle +\mathbf{B}| + \langle g_{00,-\mathbf{B}}(\mathbf{r})| \otimes \beta\ {}_B\langle -\mathbf{B}| \right) \Big] \\ &+ \frac{1}{2}\Big[ \left( |g_{00,+\mathbf{B}}(\mathbf{r})\rangle \otimes \beta |+\mathbf{B}\rangle_B - |g_{00,-\mathbf{B}}(\mathbf{r})\rangle \otimes \alpha |-\mathbf{B}\rangle_B \right)\left( \langle g_{00,+\mathbf{B}}(\mathbf{r})| \otimes \beta^*\ {}_B\langle +\mathbf{B}| - \langle g_{00,-\mathbf{B}}(\mathbf{r})| \otimes \alpha^*\ {}_B\langle -\mathbf{B}| \right) \Big]. \end{aligned} \tag{8}$$

We then trace out the spatial degrees of freedom $\left\{ |+\mathbf{B}\rangle_B, |-\mathbf{B}\rangle_B \right\}$ by calculating

$$\mathrm{Tr}_B \rho_B = \mathrm{Tr}_{\{|+\mathbf{B}\rangle,|-\mathbf{B}\rangle\}} \rho_B = {}_B\langle +\mathbf{B}|\rho_B|+\mathbf{B}\rangle_B + {}_B\langle -\mathbf{B}|\rho_B|-\mathbf{B}\rangle_B. \tag{9}$$

We note that

$$\langle +\mathbf{B}|_B\, \rho_B\, |+\mathbf{B}\rangle_B = \frac{1}{2}\Big[ \left( \alpha^* |g_{00,+\mathbf{B}}(\mathbf{r})\rangle \right)\left( \langle g_{00,+\mathbf{B}}(\mathbf{r})| \alpha \right) \Big] + \frac{1}{2}\Big[ \left( \beta |g_{00,+\mathbf{B}}(\mathbf{r})\rangle \right)\left( \langle g_{00,+\mathbf{B}}(\mathbf{r})| \beta^* \right) \Big]$$

$$\langle -\mathbf{B}|_B\, \rho_B\, |-\mathbf{B}\rangle_B = \frac{1}{2}\Big[\big(\beta^*\big|g_{00,-\mathbf{B}}(\mathbf{r})\rangle\big)\big(\langle g_{00,-\mathbf{B}}(\mathbf{r})\big|\beta\big)\Big] + \frac{1}{2}\Big[\big(\alpha\big|g_{00,-\mathbf{B}}(\mathbf{r})\rangle\big)\big(\langle g_{00,-\mathbf{B}}(\mathbf{r})\big|\alpha^*\big)\Big], \quad (10)$$

and hence we obtain the reduced density matrix that describes Bob's clocks:

$$\rho_{clocks} = \mathrm{Tr}_B \langle \rho_B \rangle =$$
$$\frac{1}{2}\Big[\big(|\alpha|^2+|\beta|^2\big)\big|g_{00,+\mathbf{B}}(\mathbf{r})\rangle \otimes \langle g_{00,+\mathbf{B}}(\mathbf{r})\big|\Big] + \frac{1}{2}\Big[\big(|\beta|^2+|\alpha|^2\big)\big|g_{00,-\mathbf{B}}(\mathbf{r})\rangle \otimes \langle g_{00,-\mathbf{B}}(\mathbf{r})\big|\Big] =$$
$$= \frac{1}{2}\big|g_{00,+\mathbf{B}}(\mathbf{r})\rangle \otimes \langle g_{00,+\mathbf{B}}(\mathbf{r})\big| + \frac{1}{2}\big|g_{00,-\mathbf{B}}(\mathbf{r})\rangle \otimes \langle g_{00,-\mathbf{B}}(\mathbf{r})\big|, \quad (11)$$

which is maximally mixed (the probabilities of detecting the original locations $|\mathbf{o}_B+\mathbf{B}\rangle_B = |+\mathbf{B}\rangle_B$, $|\mathbf{o}_B-\mathbf{B}\rangle_B = |-\mathbf{B}\rangle_B$ are both equal to ½). This matrix is completely independent of any basis used in the description of Alice's measurement process - precisely what we need to ensure that the "no signaling" principle is obeyed.

Going back to our "spin-based" gedanken experiment (from the main text), we can ask: Would the quantization of linearized gravity maintain relativistic causality, as it did for the gedanken experiment presented here? Would there be equivalent consequences to the two gedanken experiments?

We show that the answer to both questions is **no**. The spin offers unique consequences.

While it seems at first that the above spatial gedanken experiment is similar to the spin gedanken experiment, they are inherently different: The algebra of spin addition is different from that of spatial coordinates. The superposition of spin states along the $\hat{\mathbf{x}}$ axis can end up in a spin oriented along the $\hat{\mathbf{y}}$ axis, which cannot occur with spatial coordinates. In practical terms, it seems that **time dilation measurements can localize a particle but they cannot determine its spin orientation**. So it is the richness of the spin algebra, and the specific way in which it couples to the spacetime, that leads to interesting and important consequences, which cannot be obtained with the spatial version of the gedanken experiment.

*Section 3: A quantum description for the spacetime associated with a single spin-half particle and a quantum measurement with an ideal clock*

In this section, we attempt to construct a quantum description of a single spin-half particle that includes both its spin and the surrounding spacetime. We begin by describing the quantum state of a stationary (i.e. completely delocalized) spin-half particle with a specific spin state $\left|\mathbf{S}^{\mu}\right\rangle$ that is coupled to the spacetime. This state is translation invariant, and therefore translating the particle's state and summing over all the possible translations, we obtain the following quantum representation:

$$\left|\mathbf{S}^{\mu}\right\rangle \otimes \int d^{3}\mathbf{x} \exp\left(-\tfrac{i}{\hbar} E_{0} t\right)\left|ST(\mathbf{x},\mathbf{S}^{\mu})\right\rangle , \qquad (12)$$

where $E_0$ is the rest mass and $\left|ST(\mathbf{x},\mathbf{S}^{\mu})\right\rangle$ denotes the spacetime quantum ket vector associated with a spin $\mathbf{S}^{\mu}$ located at $\mathbf{x}$.

By boosting this stationary particle state, we obtain the state of a completely delocalized particle with momentum $\mathbf{p}$:

$$\left|\mathbf{S}^{\mu}(\mathbf{p})\right\rangle \otimes \int d^{3}\mathbf{x} \exp\left(\tfrac{i}{\hbar}(\mathbf{p}\cdot\mathbf{x} - E_{\mathbf{p}} t)\right) B(\mathbf{p})\left|ST(\mathbf{x},\mathbf{S}^{\mu})\right\rangle, \qquad (13)$$

where $\left|\mathbf{S}^{\mu}(\mathbf{p})\right\rangle = \left|\Lambda^{\mu}_{\mu'}(\boldsymbol{\beta}_{\mathbf{p}})\mathbf{S}^{\mu'}\right\rangle$ is the Dirac spinor associated with momentum $\mathbf{p}$, spin $\mathbf{S}^{\mu}$, boost spin transformation $\Lambda^{\mu}_{\mu'}(\boldsymbol{\beta}_{\mathbf{p}})$, velocity $\boldsymbol{\beta}_{\mathbf{p}} = \mathbf{u}_{\mathbf{p}}/c$, $(\mathbf{p} = (1-\boldsymbol{\beta}_{\mathbf{p}}{}^{2})^{-1/2} m_{0}\mathbf{u}_{\mathbf{p}})$, and a boost operator $B(\mathbf{p})$ imbuing a particle at rest with momentum $\mathbf{p}$. It should be noted that $B(\mathbf{p})$ acts on the entire spacetime of the particle, applying to it a Lorenz transformation $L^{\mu}_{\mu'}(\boldsymbol{\beta}_{\mathbf{p}})$. The spacetime metric $\tilde{g}_{\alpha\beta}\left[x^{\nu}\right]$ associated with the boosted spacetime ket state $B(\mathbf{p})\left|ST(\mathbf{x},\mathbf{S}^{\mu})\right\rangle$ is

just a Lorentz transformation $L^{\mu}_{\mu'}(\boldsymbol{\beta}_{\mathbf{p}})$ of the spacetime metric $g_{\alpha\beta}\left[x^{\nu};\left|ST(\mathbf{x},\mathbf{S}^{\mu})\right\rangle\right]$ associated with the state $\left|\text{Spacetime of spin } \mathbf{S}^{\mu} \text{ at } \mathbf{x}\right\rangle$ - i.e.,

$$\tilde{g}_{\alpha\beta}[x^{\nu}] = L_{\alpha}{}^{\alpha'}(\boldsymbol{\beta}_{\mathbf{p}})L_{\beta}{}^{\beta'}(\boldsymbol{\beta}_{\mathbf{p}})g_{\alpha'\beta'}\left[L^{\nu}{}_{\nu'}(\boldsymbol{\beta}_{\mathbf{p}})x^{\nu'};\left|ST(\mathbf{x},\mathbf{S}^{\mu})\right\rangle\right], \tag{14}$$

Finally, we construct a general quantum state as the superposition of these boosted states, and thus it is formally described by,

$$\sum_{\mathbf{S}^{\mu}}\int d^{3}\mathbf{p}\, a\left(\mathbf{p},\mathbf{S}^{\mu}\right)\left|\mathbf{S}^{\mu}(\mathbf{p})\right\rangle \otimes \int d^{3}\mathbf{x}' \exp\left(i/\hbar(\mathbf{p}\cdot\mathbf{x}-E_{\mathbf{p}}t)\right)B(\mathbf{p})\left|ST(\mathbf{x},\mathbf{S}^{\mu})\right\rangle, \tag{15}$$

where, $a\left(\mathbf{p},\mathbf{S}^{\mu}\right)$ are the amplitudes associated with each momentum and spin state. It is important to note that even-though that each plane wave solution as in (13) is completely delocalized, we are particularly interested in the case in which their superposition is localized.

Using this approach, we now turn to analyze our gedanken experiment quantum mechanically. To do this, we shall calculate the possible values of the spacetime metric element $g_{00}$ at an arbitrary spacetime point $(0,\mathbf{x}_0)$. We shall compute this with the aid of an ideal quantum clock located at time $t=0$ at $\mathbf{x}_0$ . An ideal clock could be a qubit that is used for tracking the time. It could be, for example, a two-level atomic system of the form $\left|\varphi_{\text{clock}}(t)\right\rangle = \left(\left|0\right\rangle + \exp(-i\omega_{\text{clock}}t)\left|1\right\rangle\right)\otimes\int d\mathbf{x}_c\varphi(t,\mathbf{x}_c-\mathbf{x}_0)\left|\mathbf{x}_c\right\rangle$, where $\varphi_c(t,\mathbf{x}_c-\mathbf{x}_0)$ is a wavepacket (located at the vicinity of $\mathbf{x}_0$) describing the position of the clock and $\left|0\right\rangle + \exp(-i\omega_{\text{clock}}t)\left|1\right\rangle$ is a time dependent qubit (realized, e.g., by two energy levels of the atom[95]). For simplicity, we shall now assume that the clock's mass and energy are small when compared to the mass and energy of the spin-half particle (but the following calculations can be generalized provided that the clock and the particle are separated by a sufficiently large distance). Furthermore, before we continue, let us note that the evolution of the clock's wave function under the influence of an external gravitational field is given by,

$$\left|\varphi_{\text{clock}}(dt)\right\rangle = \\ = \int d^3\mathbf{x}_c \left(|0\rangle + \exp\left(-i\theta_{\text{clock}} dt\sqrt{g_{00}[0,\mathbf{x}_c]}\right)|1\rangle\right) \otimes \varphi_c(dt\sqrt{g_{00}[0,\mathbf{x}_c]}, \mathbf{x}_c - \mathbf{x}_0)|\mathbf{x}_c\rangle. \tag{16}$$

Thus, by measuring the qubit's $|0\rangle + \exp\left(-i\theta_{\text{clock}} dt\sqrt{g_{00}[0,\mathbf{x}_0]}\right)|1\rangle$ state in the computational basis, we can find the possible values of the parameter $\sqrt{g_{00}[0,\mathbf{x}_0]}$. Now let us consider the case where the clock is influenced by a *quantum superposition of several gravitational fields* (induced by the wavefunction of the spin-½ particle $\left|\psi_{\text{particle}}\right\rangle$). This measurement is described by noting that the quantum state of the *combined system,* i.e., the electron and the clock's time pointer at time $dt$, is given by

$$\left|\psi_{\text{particle and clock's pointer}}(dt)\right\rangle = \\ = \sum_{\mathbf{S}^\mu}\int d^3\mathbf{p}\, a\left(\mathbf{p},\mathbf{S}^\mu\right)\left|\Lambda^\mu_{\ \mu'}(\mathbf{p})\mathbf{S}^{\mu'}\right\rangle \otimes \\ \otimes \int d^3\mathbf{x} \exp\left({}^{i}\!/\!_{\hbar}(\mathbf{p}\cdot\mathbf{x} - E_{\mathbf{p}}dt)\right) B(\mathbf{p})\left|ST(\mathbf{x},\mathbf{S}^\mu)\right\rangle \otimes \\ \otimes \int d^3\mathbf{x}_c \varphi_c\left(dt\sqrt{L_0^{\ \alpha}(\boldsymbol{\beta}_{\mathbf{p}})L_0^{\ \beta}(\boldsymbol{\beta}_{\mathbf{p}})g_{\alpha\beta}\left[L^{\nu}_{\ \nu'}(\boldsymbol{\beta}_{\mathbf{p}})\cdot\left(0,\mathbf{x}_c - \mathbf{x}\right);\left|ST(\mathbf{x},\mathbf{S}^{\mu})\right\rangle\right]}, \mathbf{x}_c - \mathbf{x}_0\right)|\mathbf{x}_c\rangle \otimes \\ \otimes\left(|0\rangle + \exp\left(-i\theta_{\text{clock}} dt\sqrt{L_0^{\ \alpha}(\boldsymbol{\beta}_{\mathbf{p}})L_0^{\ \beta}(\boldsymbol{\beta}_{\mathbf{p}})g_{\alpha\beta}\left[L^{\nu}_{\ \nu'}(\boldsymbol{\beta}_{\mathbf{p}})\cdot\left(0,\mathbf{x}_c - \mathbf{x}\right);\left|ST(\mathbf{x},\mathbf{S}^{\mu})\right\rangle\right]}\right)|1\rangle\right), \tag{17}$$

where, as explained above, $L_0^{\ \alpha}(\boldsymbol{\beta}_{\mathbf{p}})L_0^{\ \beta}(\boldsymbol{\beta}_{\mathbf{p}})g_{\alpha\beta}\left[L^{\nu}_{\ \nu'}(\boldsymbol{\beta}_{\mathbf{p}})\cdot\left(0,\mathbf{x}_c - \mathbf{x}\right);\left|ST(\mathbf{x},\mathbf{S}^{\mu})\right\rangle\right]$ is the time dilation effect (at $\mathbf{x}_c$) which is associated with the boosted state $B(\mathbf{p})\left|\text{Spacetime of spin } \mathbf{S}^\mu \text{ at } \mathbf{x}\right\rangle$.

Finally, to calculate the possible values of the clock's time dilation, we simply trace out all the degrees of freedom associated with the electron and the clock's position. This way we obtain the clock's density matrix:

$$
\begin{aligned}
\rho_{clcok}(dt) = &\int d^3\mathbf{x}\sum_{\mathbf{S}^\mu}\int d^3\mathbf{q}\int d^3\mathbf{p}\,\bar{a}\left(\mathbf{q},\mathbf{S}^\mu\right)a\left(\mathbf{p},\mathbf{S}^\mu\right)\exp\left(\tfrac{i}{\hbar}(\mathbf{p}-\mathbf{q})\mathbf{x}-\tfrac{i}{\hbar}(E_{\mathbf{p}}-E_{\mathbf{q}})dt\right)\cdot \\
&\cdot\int d^3\mathbf{x}_c\varphi_c\left(dt\sqrt{L_0^{\ \alpha}(\boldsymbol{\beta}_{\mathbf{p}})L_0^{\ \beta}(\boldsymbol{\beta}_{\mathbf{p}})g_{\alpha\beta}\left[L^{\nu}{}_{\nu'}(\boldsymbol{\beta}_{\mathbf{p}})\cdot\left(0,\mathbf{x}_c-\mathbf{x}\right);\left|ST\left(\mathbf{x},\mathbf{S}^\mu\right)\right\rangle\right]},\mathbf{x}_c-\mathbf{x}_0\right)\cdot \\
&\quad\cdot\varphi_c\left(dt\sqrt{L_0^{\ \alpha}(\boldsymbol{\beta}_{\mathbf{q}})L_0^{\ \beta}(\boldsymbol{\beta}_{\mathbf{q}})g_{\alpha\beta}\left[L^{\nu}{}_{\nu'}(\boldsymbol{\beta}_{\mathbf{q}})\cdot\left(0,\mathbf{x}_c-\mathbf{x}\right);\left|ST\left(\mathbf{x},\mathbf{S}^\mu\right)\right\rangle\right]},\mathbf{x}_c-\mathbf{x}_0\right)^*\cdot \\
&\cdot\left(\left|0\right\rangle+\exp\left(-i\theta_{\text{clock}}dt\sqrt{L_0^{\ \alpha}(\boldsymbol{\beta}_{\mathbf{p}})L_0^{\ \beta}(\boldsymbol{\beta}_{\mathbf{p}})g_{\alpha\beta}\left[L^{\nu}{}_{\nu'}(\boldsymbol{\beta}_{\mathbf{p}})\cdot\left(0,\mathbf{x}_c-\mathbf{x}\right);\left|ST\left(\mathbf{x},\mathbf{S}^\mu\right)\right\rangle\right]}\right)\left|1\right\rangle\right) \\
&\quad\left(\left\langle 0\right|+\exp\left(+i\theta_{\text{clock}}dt\sqrt{L_0^{\ \alpha}(\boldsymbol{\beta}_{\mathbf{q}})L_0^{\ \beta}(\boldsymbol{\beta}_{\mathbf{q}})g_{\alpha\beta}\left[L^{\nu}{}_{\nu'}(\boldsymbol{\beta}_{\mathbf{q}})\cdot\left(0,\mathbf{x}_c-\mathbf{x}\right);\left|ST\left(\mathbf{x},\mathbf{S}^\mu\right)\right\rangle\right]}\right)\left\langle 1\right|\right).
\end{aligned}
\tag{18}
$$

In particular, we see that spin-spacetime censorship is maintained if the spacetime metric is spin-independent. Note that this is a sufficient, but not necessary, solution to the question – how relativistic causality and no-cloning are maintained in our gedanken experiment.

Elaborating on the possibility that the spacetime metric is spin-independent, we note that it imposes a very strong condition of spin censorship, as it implies that the spacetime associated with a single spin is spherically symmetric. Assuming that the metric is indeed spherically symmetric can reduce the number of possible spacetime metrics according to Birkhoff's generalized theorem[91]. For example, one could consider the spherically symmetric Schwarzschild metric centered at the localized spin. This choice is consistent with our previously described dust stress-energy tensors approach (classical approach #2). However, if the spacetime (around the spin-½) is indeed spherically symmetric, then one may also expect certain classical ramifications for the EMFE. For example, many particles having all their spins pointing in the same direction, and thus forming a very strong magnetic field, may still show no break of spherical symmetry of the time dilation effect around them. Therefore, this particular solution (maintaining causality and no cloning in our gedanken experiment) seems to lead to a classical modification of EMFE with implications on cosmological scales, unless it can somehow be shown that the collective spacetime effect of many particles is different from that of a single particle (see e.g., the paragraph regarding an additive auxiliary stress energy tensor in SI section 5).

A different pathway for achieving spin censorship based on the formalism here involves an idea related to weak measurements[96]: consider a situation where the clock's state has a very broad distribution around the expectation value of $\theta_{\text{clock}}$ from which it is practically impossible to infer the axis of the spin (the back-action of the pointer on the measured system could also contribute to the accumulated uncertainty). Applying weak measurement is quite plausible due to the minuscule coupling strength between the spin and the clock, leading to a shift in the clock pointer that may be much smaller than its quantum uncertainty (even if the clock is very precise). The less obvious characteristic of this approach is finding how the measurement strength increases with duration and with the number of spins/clocks.

## *Section 4: Spin spacetime censorship for photons*

In this section, we present an analogue to the spin gedanken experiment that applies to massless particles with spin, such as photons with polarizations. Photons can be entangled through their polarizations, for example, entanglement of the vertical ($V$) and horizontal ($H$) polarization according to $\left(|HV\rangle - |VH\rangle\right)/\sqrt{2}$. To explain such a gedanken experiment, recall that according to the EMFE, photons that are linearly polarized induce (very) weak gravitational pp-waves[97] with (vertical-horizontal) polarization relative to the photon polarization state (thus linearly polarized photons at $\pm 45^0$ induce a polarized mode of gravitational pp-wave at $\pm 45^0$). It therefore follows that if Alice projects her photon to a specific linearly polarized state, then Bob can (in theory) measure the gravitational wave induced by his photon to determine the induced polarization state. From a classical point of view, by measuring gravity waves, Bob can determine whether his photon is linearly polarized in one of ($0^0$,$90^0$) linear polarization states or whether it is linearly polarized in one of the ($+45^0$,$-45^0$) states. Again, if Alice is sufficiently far, this clearly violates relativistic causality. Altogether, we expect there to be spin censorship principles for any spin, not only ½.

## *Section 5: Attempts to explain the gedanken experiment results with classical speculative approaches*

### *Additive auxiliary tensor*

One could attempt to realize a spin-spacetime censorship mechanism, by adding an auxiliary aspherical spin-dependent stress-energy tensor that would cancel the aspherical part of the Maxwell stress-energy tensor. With this additional term, the spin-spacetime censorship is realized by construction for each particle. Next, one can imagine what happens if there are many particles (each with its own additional auxiliary stress energy tensor). Naturally, one would assume that these additional auxiliary stress energy tensors add up together (yielding a total tensor which, in contrast to the Maxwell tensor, is at most linearly proportional to the number of particles). This approach seems rather ad hoc, assuming a yet unknown stress energy tensor that somehow accompanies the ordinary Maxwell stress energy tensor, but it might find justification from other perspectives.

### *Electric dipole*

Other censorship mechanisms may seem appealing at first but prove to be flawed. For example, one might suggest that the looked-for electron electric dipole[98,99] can eliminate the aspherical parts of the stress-energy tensor created by the spin magnetic dipole. Indeed, the standard model predicts a non-zero electron electric dipole moment[98,99], whose value has not been found yet. However, the state-of-the-art upper bound on the electron's electric dipole moment found experimentally[100] is too small to compensate for an aspherical spacetime curvature created by the much larger electron dipole moment.

*Electron rotation (classical spin)*

Another candidate for a censorship mechanism is based on attributing a different internal rotation rate (classical "spin") to the electron, so it bends spacetime in an aspherical way that cancels out the effect of the electron magnetic dipole moment. However, one can show that such a rotation cannot compensate for the asphericity arising due to the spatial extent of the magnetic field, unless the electron is taken to have an extended mass/charge distribution. Such an approach would corroborate past attempts to treat the electron as a Kerr-Newman black hole that has a rotation rate and a spin consistent with each other. For example, Carter showed[101] that a constant classical angular momentum of $\hbar/2$ gives rise to magnetic moment similar to that of the electron spin (see also[102]). Trying now to alter the rotation rate to ensure a spherically symmetric spacetime curvature would harm this essential consistency. When exploring these models, it is worth noting that they have been shown to suffer from a naked singularity[103] and closed timelike curves[103].

However, a relatively new ghost-free approach to gravity, known as infinite derivative gravity[104,105], can overcome this problematic ring singularity. In such a non-local theory, the gedanken experiment as a whole, and entanglement in particular, would have to be carefully analyzed before conclusions can be made.

## *Section 6: Revisiting non-commutative spacetime geometry and other quantum approaches that seem to be challenged*

In this section, we revisit the proposed censorship mechanism of non-commutative geometry of spacetime mentioned in the main text. We can introduce a generalization of our gedanken experiment, which seems to prove that this mechanism cannot maintain causality for all possible setups, of the gedanken experiment: The signaling protocol between Alice and Bob can be performed with many pairs of entangled particles (used simultaneously and measured by Alice with a single Stern-Gerlach device), while Bob measures each of his particles separately with *two clocks per particle* (so that the two clocks are positioned symmetrically at locations $\pm L\hat{\mathbf{n}}$ with respect to the spin-½ particle). Bob also locates his clocks at different $\hat{\mathbf{n}}$ orientations around each particle. Again, he can then compare the clocks times and determine that the clocks in the $\pm\hat{\mathbf{x}}$ orientation, are faster or slower than the clocks placed in the $\pm\hat{\mathbf{y}}$ axes orientation. Now since there is one clock per particle, Bob can separate his particles (sufficiently far apart) to prevent any possible interference between the different time measurements. This generalized gedanken experiment shows that mutual influence between different clocks cannot be given as a reason to resolve the paradox, which appears to challenge non-commutative geometry of spacetime as a censorship mechanism. Other possible approaches can still be *combined with a non-commutative geometrical model* in an attempt to maintain relativistic causality in our gedanken experiment.

Another approach for maintaining relativistic causality (in our gedanken experiment) could be based on coupling the wave equations of the electron to those of the gravitational potential: adding the gravitational potential (derived via the Poisson equation with the electron's mass density as a source) to the electron quantum wave equation (e.g., the Newton-Schrodinger equation and its relativistic generalizations). We did not consider this type of censorship candidates because they were shown, alongside with other nonlinear modifications of the Schrodinger equation, to allow signaling[106].

## *Section 7: The Kerr–Newman metric in classical gravity*

This section recalls the aspherical solution of the EMFE for an electron with a spin - For a comprehensive discussion of this topic see[107]. An exact electro-vacuum solution can be expressed in the Boyer–Lindquist coordinates $t, r, \theta, \phi$ ($x = \sqrt{r^2 + a^2}\sin\theta\cos\phi,\ y = \sqrt{r^2 + a^2}\sin\theta\sin\phi,\ z = r\cos\theta$) with the spacetime metric

$$g_{\mu\nu} = \begin{bmatrix} c^2(\Delta - a^2\sin^2\theta)/\rho^2 & 0 & 0 & -2ac\sin^2\theta(\Delta - r^2 - a^2)/\rho^2 \\ 0 & -\rho^2/\Delta & 0 & 0 \\ 0 & 0 & -\rho^2 & 0 \\ -2ac\sin^2\theta(\Delta - r^2 - a^2)/\rho^2 & 0 & 0 & \sin^2\theta(a^2\Delta\sin^2\theta - r^4 - 2r^2a^2 - a^4)/\rho^2 \end{bmatrix}, \tag{16}$$

where $a = \dfrac{\hbar}{2Mc}$, $\rho^2 = r^2 + a^2\cos^2\theta$, $\Delta = r^2 - r_s r + a^2 + r_Q^2$, $r_s = \dfrac{2Gm_e}{c^2}$, $r_Q^2 = \dfrac{e^2 G}{4\pi\epsilon_0 c^4}$, and the electromagnetic vector potential is given by[107,108] $A_\mu = \left(\dfrac{r\, r_Q}{\rho^2}, 0, 0, -\dfrac{c^2\, a\, r\, r_Q \sin^2\theta}{\rho^2\, G\, m_e}\right)$. It should be noted that for all known spin-1/2 particles the Kerr–Newman metric leads to a naked singularity and to closed time-like curves109. The radius of this ring singularity is $r_a \approx \dfrac{\hbar}{2mc}$, which is 1.93×10$^{-13}$[m] for an electron. To avoid this ring singularity and other UV problems, Biswas et al.[110] have suggested a higher derivative covariant generalization of general relativity, which could be applied to our gedanken experiment as well. Other researchers have suggested different models for spin-1/2 particles (see e.g.,[111]).

Another aspect of the Kerr-Newmann spacetime is that it predicts the electric dipole moment of the electron to be zero (the expectations for a nonzero electron dipole moment are related to QED). However, we know that higher, extremely tiny, but non-zero even-multipoles are expected to exist in Kerr-Newmann spacetime. No such multipoles have been observed so far in experiments. There is a rich discussion in the literature about suitable modifications and perhaps an internal structure[111-118] that could make these multipoles much smaller, so to become consistent with current experiments. In either case, the higher multipoles would not prevent the break of the spacetime spherical symmetry at the heart of our gedanken experiment. Consequently, our work show that even generalized version of the Kerr-Newmann spacetime cannot model our gedanken experiment without a causality paradox. It is intriguing that quantum information considerations have such consequences on a basic problem in (classical) general relativity.

## *Section 8: Perturbative approach with linearized quantum gravity*

In this section, we examine the gedanken experiment from the perspective of linearized quantum gravity (two recent accounts appear in[119,120]), where gravitons emerge through second quantization of a linearized perturbation to the metric. The inherently relativistic dynamics of these second-quantized linearized gravitational fields may seem to automatically solve the issues of causality (and retardation) in a manner similar to the case of quantum electrodynamics (QED). That is, in the same way that QED prevents measuring the spin with photons, linearized quantum gravity could be expected to prevent measuring the spin with gravitons.

In this approach, we analyze the spin coupling to spacetime by considering a *linearized theory*, where basic concepts that are familiar from flat space, such as angular momentum and dipole moment, carry over to curved spacetime[121].These concepts are needed to model our gedanken experiment. Therefore, one would naturally raise the question of whether linearized quantum gravity could model our gedanken experiment with no causality paradox.

To avoid the paradox, a specific mechanism is needed in linearized quantum gravity: one that will prevent Bob from inferring the axis and direction of the spin with his clocks. However, a few difficulties are encountered.

(1) There seem to be some core differences between QED and linearized quantum gravity (see also Methods section 2), related to the different ways these theories couple to spin. Within the former, the scattering amplitude typically depends on the spin of the fermions (i.e. before taking the customary average of initial spins and sum of the final spins). In contrast, within the latter, it is unclear how to model the scattering by a spin (e.g., would scattering off the gravitational potential created by the spin depend on the entire state of the spin, or only its axis? and if so then how?).

(2) The literature seems to mostly analyze scenarios where the sources of the fields are classical[119,120], while the non-commutativity of spin components in our gedanken experiment makes it inherently quantum. Therefore, a quantum theory of linearized gravity *with non-classical sources* is needed.

(3) It is possible to use the ADM formalism and the Wheeler-DeWitt equation in its linearized form to describe the way in which spin sources gravity. However, in Methods section 4 we proved that this approach fails (unless we assume that spin and spacetime are completely decoupled).

(4) Even in the case of a theory that treats a non-classical source, it seems that for preserving the total angular momentum, simple diagrams would not suffice: The graviton has spin 2, while the electron has spin ½ and thus coupling them seems to invoke multiple mediators (both photons and gravitons). Such a multi-particle interaction seems to require nonlinear interactions (i.e. vertices involving an electron, a photon and a graviton), which go beyond the scope of linearized quantum gravity. Alternatively, we can use higher order diagrams (see Fig. 4a) that are, however, diverging (containing at least one loop). There are other possibilities (e.g., Fig. 4b and its inverse) that include processes such as photon emission by the spin absorbing a graviton (could be envisioned as radiation from a freely-falling electron). However, as of now, such processes are at most speculative.

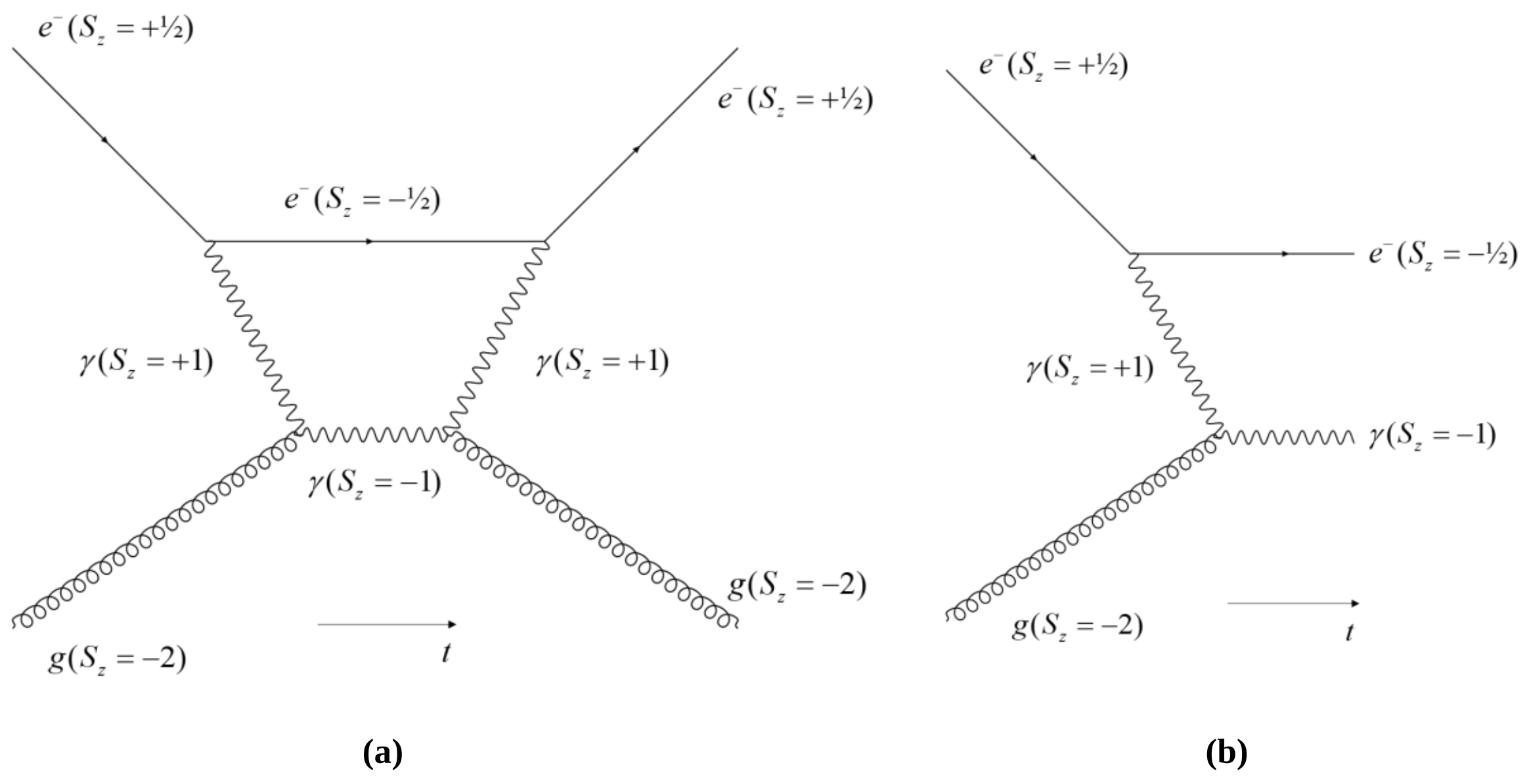

**Figure 4: Feynman diagrams in linearized quantum gravity.** (a) High order diagram containing one loop. The electron maintains its spin. (b) The electron flips its spin as a result of the interaction, in which it absorbs a graviton and emits a photon (possibly describing a spin-flip radiation process by fee-falling in the gravitational field).

To conclude, we are not aware of any satisfactory mechanism addressing our gedanken experiment within the current theory of linearized quantum gravity.

## *Section 9: Why a single spin-½ cannot be fully inferred with magnetic measurements*

This section elaborates on the question of why the spin axis cannot be determined by measuring its induced magnetic field. It is the back-action of the quantum spin measurements on the spin axis that affects the spin and changes it (unless the measurements are performed along the spin axis). It seems plausible that a similar back-action would enable modeling our gedanken experiment without a contradiction with causality. We elaborate on this possibility in Methods section 4 and SI section 8, showing mechanisms of back-action in quantum gravity that still result in a contradiction, unlike the case of quantum electrodynamics that we describe in this section.

A measuring apparatus (made of coils, magnets, etc.) that measures a spin does so by coupling to the spin – applying on it a magnetic field $\mathbf{B} = B\hat{\mathbf{n}} \neq 0$ even when spatially separated from the spin. The spin-½ state is influenced by this magnetic field. The magnetic moment $\mu$ of the spin and the magnetic field $\mathbf{B} = B\hat{\mathbf{n}}$ determine the spin's time dependent unitary evolution, which is given by the Schrödinger equation according to $i\hbar\partial_t \left|S(t)\right\rangle = \left( \frac{\mu B}{2} \left|+\hat{\mathbf{n}}\right\rangle\left\langle+\hat{\mathbf{n}}\right| - \frac{\mu B}{2} \left|-\hat{\mathbf{n}}\right\rangle\left\langle-\hat{\mathbf{n}}\right| \right)\left|S(t)\right\rangle$. The spin-½ evolution is thus given by $\left|S(t)\right\rangle = \gamma_{+\hat{\mathbf{n}}} \exp\left( i\frac{\mu B}{2\hbar}t \right)\left|+\hat{\mathbf{n}}\right\rangle + \gamma_{-\mathbf{n}} \exp\left( -i\frac{\mu B}{2\hbar}t \right)\left|-\hat{\mathbf{n}}\right\rangle$, where $\gamma_{\pm\hat{\mathbf{n}}}$ are constants that depend on initial conditions. Eventually, the state of the system undergoes a decoherence process when coupled to macroscopic measuring apparatus, so that its final state is described by the density matrix $\rho = p_{+\hat{\mathbf{n}}} \left|+\hat{\mathbf{n}}\right\rangle\left\langle+\hat{\mathbf{n}}\right| + p_{-\mathbf{n}} \left|-\hat{\mathbf{n}}\right\rangle\left\langle-\hat{\mathbf{n}}\right|$, with $p_{\pm\hat{\mathbf{n}}} = \left|\gamma_{\pm\hat{\mathbf{n}}}\right|^2$ denoting the probability for the measurement outcome of $\left|+\hat{\mathbf{n}}\right\rangle$ and $\left|-\hat{\mathbf{n}}\right\rangle$, respectively. This way, the field axis $\mathbf{B} = B\hat{\mathbf{n}}$ determines the possible outcomes of the spin-½ measurement process (either $\left|+\hat{\mathbf{n}}\right\rangle$ or $\left|-\hat{\mathbf{n}}\right\rangle$). Formally, we denote the operator describing the measurement process by $S_{+\hat{\mathbf{n}}} = \left|+\hat{\mathbf{n}}\right\rangle\left\langle+\hat{\mathbf{n}}\right| - \left|-\hat{\mathbf{n}}\right\rangle\left\langle-\hat{\mathbf{n}}\right|$. This operator is subjected to the well-known commutation relations: $[\mathbf{S_x}, \mathbf{S_y}] = i\mathbf{S_z}$, $[\mathbf{S_y}, \mathbf{S_z}] = i\mathbf{S_x}$, $[\mathbf{S_z}, \mathbf{S_x}] = i\mathbf{S_y}$, as well as to the restriction on inferring simultaneously the components $\mathbf{S_x}, \mathbf{S_y}, \mathbf{S_z}$ of the spin (or equivalently the impossibility of inferring simultaneously all the components of $\mathbf{B}$).

*Section 10: Measurement of the entire spin state without altering the spin*

We have seen that under certain candidate theories of gravity, Bob can measure the axis of the spin without altering it. This capability seems to lead to a causality paradox, even without finding the entire state of the spin. In this section, we show a simple equivalence between measuring the spin axis and measuring the entire spin state. In other words, we show that any possibility to measure the spin axis would automatically enable to measure the spin direction, finding the only missing bit of information beyond the axis without altering the spin state. The direct implication is that finding the spin axis leads to a paradox that is equivalent to violation of the "no-cloning" theorem.

The idea in short is that once Bob finds the axis of the spin, even without its direction, he can place Stern-Gerlach magnets oriented along this axis, and thus find whether the spin is up or down without altering its state. Such a Stern-Gerlach test is only possible when Bob knows in advance what the axis is. This needed advanced knowledge is why finding the spin axis through the clocks leads to the paradox with no-cloning: Bob finds the entire spin state in two steps – finding the axis with clocks and the direction with Stern-Gerlach magnets.

We present this idea in more details with the following explanation. Any state of the form $\alpha|+z\rangle+\beta|-z\rangle$ can be expressed as a point on the Bloch sphere having the form $\cos\left(\theta/2\right)|+z\rangle+\exp(i\varphi)\sin\left(\theta/2\right)|-z\rangle$ and therefore, there always exists an axis $\hat{\mathbf{n}}$ to which the state is parallel $\alpha|+z\rangle+\beta|-z\rangle\propto|+\hat{\mathbf{n}}\rangle$, i.e., proportional up to a phase to $|+\hat{\mathbf{n}}\rangle$, where $n_x=\sin(\theta)\cos(\varphi),\ n_y=\sin(\theta)\sin(\varphi),\ n_z=\cos(\theta)$. Now, by symmetrically placing many clocks on a sphere Bob can find out that the axis of the spin up to a sign so he knows that it is either $|+\hat{\mathbf{n}}\rangle$ or $|-\hat{\mathbf{n}}\rangle$ (as in Methods section 4). Finally, to find out the sign of the spin, Bob uses Stern Gerlach magnets so that the measurement axis will be parallel to the $\pm\hat{\mathbf{n}}$ axis. This

way, the Stern-Gerlach measurement does not alter the spin state. By determining this last bit of information, whether the state of the spin is $|+\hat{\mathbf{n}}\rangle$ or $|-\hat{\mathbf{n}}\rangle$, Bob has the entire state of the spin (values of $\theta$ and $\varphi$ as well as the values of $\beta / \alpha = \exp(i\varphi)\tan\left(\theta/2\right)$ ).

**Acknowledgements** We thank Y. Aharonov, R. Bekenstein, L. Diosi, J. D. Joannopoulos, A. Ori, S. Popescu, C. Roques-Carmes, M. Segev, M. Soljačić and R. Wald for helpful comments and discussions. E.C. was supported by the Canada Research Chairs (CRC) Program and by the Faculty of Engineering in Bar Ilan University. I.K. is an Azrieli Fellow, supported by the Azrieli Foundation, and was also partially supported by the Seventh Framework Programme of the European Research Council (FP7-Marie Curie IOF) under grant no. 328853-MC-BSiCS.